\documentclass[aps,pra,onecolumn,superscriptaddress,nofootinbib]{revtex4-1}

\usepackage{graphicx}
\usepackage[caption=false]{subfig}
\usepackage{latexsym}
\usepackage{amsmath}
\usepackage{amssymb}
\usepackage{amsfonts}
\usepackage{color}
\usepackage{pgf}
\usepackage{amsthm}
\usepackage{dsfont}
\usepackage{tikz}
\usetikzlibrary{matrix}
\usepackage{cancel}

\usepackage[utf8]{inputenc}
\usepackage[english]{babel}
\renewcommand{\selectlanguage}[1]{}
 
\usepackage{ulem}
\usepackage{mwe} 
\usepackage{graphicx}
\usepackage{subfig}
\usepackage{mathtools}
\usepackage{amssymb}
\usepackage{colortbl}

\usepackage{hyperref}
\newcommand{\rd}{\mathrm d}
\newcommand{\R}{\mathbb R}

\usepackage{color}
\newcommand{\C}{\mathbb C}

\newcommand{\e}{{\mathrm e}}
\newcommand{\Ccal}{{\mathcal C}}

\newcommand{\Ncal}{{\mathcal N}}

\newcommand{\rhoG}{\hat{\rho}_{\mathrm{G}}}
\newcommand{\rhoGgamma}{\hat{\rho}_{\textrm{G}, \gamma}}

\newcommand{\chiG}{\chi_{\mathrm{G}}}
\newcommand{\WG}{W_{\mathrm G}}
\newcommand{\WGgamma}{W_{\mathrm{G},\gamma}}

\newcommand{\vertiii}[1]{{\left\vert\kern-0.25ex\left\vert\kern-0.25ex\left\vert #1 
		\right\vert\kern-0.25ex\right\vert\kern-0.25ex\right\vert}}

\newcommand{\eq}[1]{\begin{equation} #1 \end{equation}}
\newcommand{\eqstar}[1]{\begin{equation*} #1 \end{equation*}}
\newcommand{\tr}{\mathrm{Tr}}
\newcommand{\Tr}{\mathrm{Tr}}

\newcommand{\eqarray}[1]{\begin{eqnarray} #1 \end{eqnarray}}

\newcommand{\ket}[1]{\left\vert #1 \right\rangle}
\newcommand{\bra} [1] {\left\langle #1 \right\vert}

\newcommand{\mean}[1]{\langle #1 \rangle}

\newcommand{\cov}[1]{\text{Cov}[#1]}
\newcommand{\rhoh}{\hat{\rho}}

\newcommand{\QCStwoG}{{\mathcal C}^2_{\mathrm G}}

\newcommand{\norm}[1]{|#1|}
\newcommand{\opt}{\mathrm{opt}}

\usepackage{graphicx}
\usepackage{dcolumn}
\usepackage{bm}
\usepackage{color}
\usepackage{soul}
\usepackage{amsmath,amsthm}
\usepackage{subfig}
\usepackage{mathtools}
\usepackage{dsfont}
\usepackage{amssymb}
\usepackage{enumitem}
\usepackage{ulem}

\begin{document}

\newcommand\dmean[1]{\langle \langle #1 \rangle \rangle}
\newcommand\hatd[1]{\hat{#1}^{\dagger}}
\newcommand\hatdn[2]{\hat{#1}^{\dagger #2}}
\newcommand\hatn[2]{\hat{#1}^{#2}}
\newcommand\matthree[9]{\left(\begin{array}{c c c} #1 & #2 & #3 \\ #4 & #5 & #6 \\ #7 & #8 & #9\end{array}\right)}
\newcommand{\UBS}{\hat{U}_{\mathrm{BS}}}

\newcommand{\Grho}{\hat{\rho}_{\mathrm{G}}}

\newtheoremstyle{break}
  {\topsep}{\topsep}%
  {\itshape}{}%
  {\bfseries}{}%
  {\newline}{}%
\theoremstyle{break}
\newtheorem{theorem}{Theorem}
\newtheorem{lem}{Lemma}
\newtheorem{rem}{Remark}
\newtheorem{defin}{Definition}
\newtheorem{corollary}{Corollary}
 \newtheorem{conj}{Conjecture}
 \newtheorem*{prop}{Properties}

\title{Tradeoff between Wigner negativity and decoherence time for cubic Gaussian states}

\author{Matthieu Arnhem}
\affiliation{Univ. Lille, CNRS, UMR 8524, INRIA - Laboratoire Paul Painlev\'e, F-59000 Lille, France}
\author{Valerio Crescimanna}
\affiliation{Department of Physics, University of Ottawa, 25 Templeton Street, Ottawa, Ontario, Canada K1N 6N5}
\author{Giuseppe Patera}
\affiliation{Univ. Lille, CNRS, UMR 8523, Laboratoire PhLAM, F-59000 Lille, France}
\author{Stephan De Bi\`evre}
\affiliation{Univ. Lille, CNRS, UMR 8524, INRIA - Laboratoire Paul Painlev\'e, F-59000 Lille, France}
\begin{abstract}

Wigner negativity has been shown to be a resource in quantum computing, raising the question of how to efficiently produce states with a large Wigner negativity. Cubic Gaussian states, obtained when a cubic gate is applied to a Gaussian state, have been proposed as candidates for this purpose.  We show that cubic Gaussian states with a  large Wigner negativity $\mathcal N$ necessarily have a short decoherence time $\tau$ that is bounded above by $\mathcal N^{-2}$. In other words, a large Wigner negativity comes at a cost: the decoherence time $\tau$ of cubic Gaussian states decreases at least quadratically in the negativity $\mathcal N$.  Maximizing $\tau$ at fixed Wigner negativity $\mathcal N$, we show that this upper bound is reached by optimal cubic Gaussian states.  They have the property that for large negativity, the optimal squeezing scales as $\ln\mathcal N$  and the optimal cubicity as $\mathcal N^{-1}$. Optimal cubic Gaussian states with large negativity can therefore be constructed with small cubicity, provided sufficient squeezing is applied. However, their decoherence time decreases with growing negativity. In addition, we show that their preparation is extremely sensitive to thermal noise.    Finally, we compare, for each integer $n$, the optimal cubic Gaussian state with the $n$th Fock state that has the same negativity: despite their pronounced differences in photon number distribution and Wigner function, we show that they have similar, be it slightly larger, decoherence times and entanglement generating potential through a beam splitter.

\end{abstract}

\maketitle

\tableofcontents

\section{Introduction}\label{sec:intro}

Wigner negativity is known to be a resource in quantum information science \cite{albarelli_resource_2018}.  
The presence of negative values of the Wigner function at some points in the phase space has, for example, been shown to be a necessary condition for quantum advantage in quantum computing protocols~\cite{mari_positive_2012, veitch_efficient_2013, pashayan_estimating_2015}. 
In particular, quantum protocols that do not use states or operations with nonpositive Wigner functions can be simulated efficiently classically \cite{mari_positive_2012, veitch_efficient_2013}. This remains true even if the Wigner functions involved are nonpositive, but their Wigner negativity is not too large \cite{pashayan_estimating_2015}.  
As a result, in quantum optics, protocols using only Gaussian states and operations can be efficiently simulated classically, whence the need to go beyond Gaussian quantum optics to obtain a quantum advantage in quantum computing \cite{Lloyd_quantum_1999}.

It has been shown furthermore that adding a cubic gate to the set of Gaussian gates does lead to a universal gate set \cite{gottesman_encoding_2001}. One way to implement such a gate is to use an ancilla prepared in the so-called ideal cubic phase state. The latter is however  unphysical, since it is nonnormalizable, has infinite mean photon number, and exhibits infinite Wigner negativity. It therefore needs to be approximated by physically realizable finite-energy states with relevant non-Gaussian and nonclassical characteristic features similar to those of the ideal cubic phase state. Cubic squeezed states have been proposed in this context \cite{gu_quantum_2009}. 
Since then, numerous efforts have been deployed in order to propose experimentally accessible protocols to generate cubic squeezed states \cite{ghose_non-gaussian_2007,marek_deterministic_2011,yukawa_emulating_2013,park_nonlinear_2014,miyata_implementation_2016,arzani_polynomial_2017,marek_general_2018,brunelli_linear_2019,hillmann_universal_2020,zheng_gaussian_2021,hahn_deterministic_2022}. 
More recently, 
experimental realizations of such states have been reported \cite{zheng_gaussian_2021,kudra_robust_2022,eriksson_universal_2023,yanagimoto_engineering_2020,kudra_robust_2022}. 

We propose here a thorough analysis of the nonclassical properties of the full six-parameter family of single mode (pure and mixed) cubic Gaussian states $\rhoGgamma=\Gamma_\gamma\rhoG\Gamma_\gamma^\dagger$, parametrized by the covariance matrix $V$ and displacement vector $\boldsymbol{\hat r}$ of the Gaussian state $\rhoG$, and by the cubicity $\gamma$ of the cubic gate $\Gamma_\gamma=\exp(-i\gamma \hat q^3)$ (see Eq.(\ref{eq:cubicgate})). To avoid confusion, let us point out that the cubic Gaussian states are, by construction, genuinely non-Gaussian states. 

Our main result relates the Wigner negativity $\mathcal N$ of the cubic Gaussian states $\rhoGgamma$ quantitatively to their decoherence time $\tau$ via the quadrature coherent scale (QCS) (see Eq.~\eqref{eq:QCSdef}) \cite{hertz_quadrature_2020}. We show that, for all cubic Gaussian states, one has that $\tau\lesssim \mathcal N^{-2}$. In other words, a large Wigner negativity entails automatically a fast decoherence of the state so that a short decoherence time is a price to pay for a large Wigner negativity.  We show furthermore that the above upper bound on the decoherence time $\tau_{\mathrm{opt}, \mathcal N}$ at fixed Wigner negativity is reached for optimal cubic Gaussian states $\hat\rho_{\mathrm{opt},\mathcal N}$,  so that 
\begin{equation}
\tau_{\mathrm{opt},\mathcal N}\sim \mathcal N^{-2}.
\end{equation}
For Wigner negativities in the range of interest, these optimal cubic Gaussian states are cubic squeezed states; their squeezing parameter $r_{\mathrm{opt}}$  increases slowly, as $\ln\mathcal N$, and their cubicity $\gamma_{\mathrm{opt}}$ decreases as $\mathcal N^{-1}$. In other words, a larger negativity with the longest possible decoherence time is achieved through a coordinated trade-off between a weaker cubicity and a stronger squeezing.
In addition, we show that the optimal cubic Gaussian states so constructed are very sensitive to thermal noise. For example, if the cubic gate is applied to a squeezed thermal state with mean photon number as small as $0.5$, only half of the negativity is obtained compared to that gained from an optimal cubic squeezed vacuum. 
 
 Our analysis uncovers in addition several further features of the optimal cubic Gaussian states. We show that the small optimal cubicity, in addition to being responsible for the increasingly large negativity of the optimal cubic Gaussian states, also  boosts the mean photon number of the squeezed vacuum by 60\%.  
Next, we show that the optimal cubic Gaussian states are clearly very different from Fock states in many ways. We will see that their Wigner functions rather than showing azimuthal symmetry,  are strongly phase dependent, with parabolic structures, and that their negativity stretches out to infinity in phase space, rather than being localized in a finite disk around the origin, as is the case for the Wigner functions of Fock states. For large negativities, their photon number probability distribution is super-Poissonian and shows a characteristic and very strong superbunching,  with a variance to mean ratio $\Delta a^\dagger a/\langle a^\dagger a\rangle\simeq 2.47$, which is about 70\% larger than the value $\Delta a^\dagger a/\langle a^\dagger a\rangle\simeq \sqrt 2$ typical for squeezed states at large squeezing. Nevertheless the optimal cubic Gaussian states and the Fock states  have also some very similar features. 
In particular, the trade-off between a short decoherence time and a large Wigner negativity that we showed to exist for cubic Gaussian states, also occurs for Fock states $|n\rangle\langle n|$. Their negativity $\mathcal N_n$ behaves as $\sqrt n$~\cite{kenfack_negativity_2004} and their decoherence time $\tau_n$ as $n^{-1}$~\cite{hertz_quadrature_2020} so that one finds $\tau_n\sim \mathcal N^{-2}_n$ here as well.
This observation suggests to calibrate the Wigner negativities of the optimal cubic Gaussian states by those of the Fock states, allowing to compare further physical features of optimal cubic Gaussian states with those of the Fock states with the same Wigner negativity (see Fig. \ref{fig:cubicphaseWignerFct1}).  To that effect, we consider, for each integer $n$, the optimal cubic Gaussian state $\hat\rho_{\mathrm{opt},\mathcal N_n}$  that has the negativity of the Fock state $|n\rangle$.  First, we find that these states have  a mean photon number close to $n$ and a QCS somewhat smaller than the QCS of $|n\rangle$. We conclude that, in this sense, optimal cubic Gaussian states provide the same Wigner negativity per photon as Fock states, but they achieve this with a somewhat longer decoherence time.  Second, it is well known that nonclassical states -- when fed into a balanced beam splitter --  produce an entangled two-mode state.  We show that the optimal cubic Gaussian states produce a near optimal amount of entanglement in this manner and slightly outperform Fock states in this protocol. Finally, we conclude our study of the optimal cubic Gaussian states by evaluating the extent to which they approximate the cubic phase states.

The various large effects from the apparently small perturbation by the cubic gate that we observe can be understood qualitatively as follows. Since, for a given Wigner negativity $\mathcal N$, a  squeezing in the $\hat p$ quadrature is applied ($r_{\mathrm{opt}}\sim \ln(\mathcal N)$), the resulting squeezed vacuum state shows, for large $\mathcal N$,  a large uncertainty in the $\hat q$ quadrature. As a result, the fast oscillations induced by the cubic gate have a large effect, even at relatively small cubicity $\gamma_{\mathrm{opt}}$. Our results give a quantitative content to this qualitative picture.

The paper is organised as follows. In Section~\ref{sec:setup} we provide the necessary background on Gaussian states and their Wigner functions. In Section~\ref{sec:cubic}-\ref{sec:wignerneg} we introduce the Wigner functions of the cubic Gaussian states and study the properties of their Wigner negativity as a function of the parameters defining the states. In Section~\ref{sec:qcs} we introduce the quadrature coherence scale of the cubic Gaussian states and recall its link with their decoherence time. Section~\ref{s:optimalCGstates} presents the optimal cubic Gaussian states. In Section~\ref{sec:comparingtoFock} we show that the optimal cubic Gaussian states with the same Wigner negativity as the $n$th Fock state have a slightly longer decoherence time and a comparable mean photon number and that they generate slightly more entanglement when used as inputs on a beam splitter.    In Section~\ref{sec:nonclass_photondistr} we study the photon number distribution of cubic squeezed vacua by analyzing analytical expressions of suitably chosen low order moments of the creation and annihilation operators that witness various nonclassical features of these distributions. Section~\ref{sec:thermal} analyzes the sensitivity of the optimal cubic Gaussian states to thermal noise. Finally, as a further application of our findings, we investigate in Section~\ref{sec:approx_ideal_cubic} how well the optimal cubic Gaussian states approach the ideal cubic state.



\section{Setup}\label{sec:setup}
We briefly introduce  those elements of the phase space formulation of quantum optics that we shall need. More details can be found, for example, in \cite{leonhardt_essential_nodate,serafini_quantum_2017,weedbrook_gaussian_2012}. 
We consider a single-mode continuous-variable system with annihilation/creation operators $\hat{a}$ and $\hatd{a}$ verifying the commutation relation $[\hat{a},\hatd{a}]=1$. The associated vector of quadratures $\boldsymbol{\hat r}=(\hat{q},\hat{p})$ is
	\begin{equation*}
	\hat{q}=\frac1{\sqrt{2}}(\hat{a}+\hatd{a}),\quad \hat{p}=-\frac{i}{\sqrt{2}}(\hat{a}-\hatd{a})\quad
	\end{equation*}
Each quantum state $\rhoh$ is determined by its characteristic function, defined as 
\begin{equation}\label{eq:charfunct}
\chi({z})=\tr \left( \rhoh \hat{D}({z}) \right), \quad z = z_1 + i z_2 = \frac1{\sqrt{2}} (a + i b), \quad{\textrm {where}}\quad
 	\hat{D}(z)=e^{z \hat a^\dag  -  z^*\hat a}
\end{equation}
is the displacement operator.
One easily checks that
\begin{equation}\label{eq:charfunct_kernel}
\chi({z})
 =\int \rd x \exp(i b x) \bra{x -\frac{a}{2}}  \rho  \ket{x + \frac{a}{2}}
 = \int \rd p' \exp( - i a p') \bra{p' + \frac{b}{2}}  \rho  \ket{p' - \frac{b}{2}},
\end{equation}
where $|x\rangle$, respectively $|p\rangle$ denote the (generalized) eigenvectors of $\hat q$, respectively $\hat p$.
The Wigner function $W$ of $\rhoh$ is defined as the Fourier transform of its characteristic function: 
	\eq{W({\alpha})=\frac{1}{\pi^2}\int\chi({z})e^{( z^* \alpha-z\cdot{\alpha}^*)}\rd^{2}{z}\label{FourierWigner}}
where $\rd^{2} \boldsymbol z=\rd\text{Re}( z)\rd\text{Im}( z) = 1/2 \rd a \rd b$ and $\alpha=\alpha_{1}+i\alpha_{2}=\frac1{\sqrt{2}} (q+ip)\in\C$. 
	It is normalized so that 
 $$
 \int W(\alpha) \rd^{2}{\alpha}=\chi(0)=\Tr\rhoh=1.
 $$
Note that, in view of what precedes,  the kernel of a state $\rhoh$ in the $\hat q$- and in the $\hat p$-basis can be expressed in terms of the Wigner function as follows:
\begin{equation}\label{eq:kernelsW}
\bra{x} \rhoh \ket{x'}=\frac12\int \exp\left(-i\left(x'-x\right)p\right)W\left(\frac{x+x'}{2},p\right)\rd p,\quad
\bra{p} \rhoh \ket{p'}=\frac12\int \exp(-i(p'-p)) W\left(q, \frac{p+p'}{2}\right)\rd q.
\end{equation}
The inverse formulas are:
\begin{equation}\label{eq:Winvkernel}
W(q,p)=  \frac{1}{\pi} \int \rd a \exp{(i a p)} \bra{q - \frac{a}{2}} \rhoh \ket{q+\frac{a}{2}}, \quad
W(q,p)=  \frac{1}{\pi} \int \rd b \exp{(- i b q)} \bra{p + \frac{b}{2}} \rhoh \ket{p - \frac{b}{2}}.
\end{equation}
 The first-order moments of a state $\rhoh$ constitute the displacement vector, defined as $\boldsymbol{r_0}=\mean{\boldsymbol{\hat r}}=\tr (\boldsymbol{\hat r}\rho)$, while the second moments make up the covariance matrix $V$ whose elements are given by
	\begin{equation}
	V_{ij}= 2 ~ \cov{\hat r_i,\hat r_j}=\mean{\{\hat r_i,\hat r_j\}}-2\mean{\hat r_i}\mean{\hat r_j}
\label{covmat}	\end{equation}
	where $\{\cdot,\cdot\}$ denotes the anticommutator.  
We will also use the notation
 \begin{equation}
 V=2\begin{pmatrix}
     \sigma_{qq}&\sigma_{qp}\\
     \sigma_{qp}&\sigma_{pp}
 \end{pmatrix}=2\Sigma.\quad
\end{equation}
With this convention, the covariance matrix $V$ of the vacuum state is $V = \mathds{1}$ and $\sigma_{qq} = \sigma_{pp} = 1/2$.

For later purposes, we introduce the squeezing operator
\begin{equation}\label{eq:squeezeop}
\hat S_\zeta=\exp\left(\frac{1}{2}\left( \zeta^*  \hatn{a}{2} - \zeta \hatdn{a}{2}\right)\right), \quad \zeta=r\exp(i \phi).
\end{equation}
One then has
\begin{equation}
\hat S_\zeta = \hat{S}_{r\exp( i \phi)}= \hat R_{\phi/2}^\dagger \hat S_r \hat R_{\phi/2},\quad \hat R_\phi=\exp(-i\phi \hat n),\quad \hat n=\hatd{a} \hat{a}.
\end{equation}
If $W$ is the Wigner function of $\rhoh$ and if $\rhoh'= \hat S_\zeta \rhoh \hat S_\zeta^\dagger$, then
\begin{equation}
    W'(q,p)=W(R_{-\phi/2}S_rR_{\phi/2}(q,p)^T), \quad\textrm{with}
    \quad S_r= \begin{pmatrix}
        e^{-r} & 0 \\
        0 & e^{r} \\
    \end{pmatrix}
    ,\quad R_{\phi}= \begin{pmatrix}
        \cos{\phi} & \sin{\phi} \\
       - \sin{\phi} & \cos{\phi} \\
    \end{pmatrix}. 
\end{equation}
A \textit{Gaussian} state $\rhoG$ is characterized by the fact that its characteristic function $\chiG$ is a Gaussian. It takes the form 
	\eqarray{\label{chiGaussian}\chiG(z)&=&e^{-\frac12\boldsymbol{z}^T\Omega V \Omega^T\boldsymbol{z}-i \sqrt{2}(\Omega {\boldsymbol{r_0}})^T\boldsymbol{z}}, \quad \Omega=\begin{pmatrix}0&1\\-1&0	\end{pmatrix}.} 
Here, $\boldsymbol{z}^T=~(z_{1}, z_{2})\in\R^{2}$.
 The Wigner function $\WG(\alpha)$ of a Gaussian state is also a Gaussian and reads
 \begin{equation}\label{eq:WignerGaussian}
     \WG(q,p)=\frac{2}{\pi\sqrt{\det V}}\exp\left(-(\mathbf{r} - \mathbf{r_0})^T V^{-1}(\mathbf{r} - \mathbf{r_0})\right),
 \end{equation}
where $\mathbf{r}^T = (q, p)$ and $\mathbf{r_0}^T = (q_0, p_0)$. 
We recall that the purity $\mathcal P_{\textrm G}$ of a Gaussian state has a simple expression in terms of its covariance matrix:
\begin{equation}
    \mathcal P_{\textrm G}=\frac1{\sqrt{\det V}}.
\end{equation}
Let $\alpha_0=\frac1{\sqrt2}(q_0+ip_0)$, then $\hatd{D}(\alpha_0)\widehat\rho_{\textrm G} \hat{D}(\alpha_0)$ is the centered Gaussian with the same covariance matrix as $\rhoG$.

The squeezed thermal (SqTh) states are the three-parameter family of Gaussian states defined as 
\begin{equation}\label{eq:SqTh}
\rhoh_{\text{SqTh}}= \hat S_\zeta \rhoh_{\text{Th}}  \hat S^\dag_\zeta \ \textrm{with}\  
\rhoh_{\text{Th}}=(1-q)\sum_n q^n\ket{n}\bra{n}.
\end{equation}
The $\rhoh_{\text{Th}}$ are referred to as thermal states, with inverse temperature $\beta$ related to $q\in[0,1[$ by  $q=e^{-\beta}$; $q$ is also related to the mean photon number $\mean{n}$ as $q=\frac{\mean{n}}{1+\mean{n}}$. 
The covariance matrix of these states is, for $\zeta=r$,
\eqstar{V_{\text{SqTh}}=\frac{1+q}{1-q}\begin{psmallmatrix}e^{-2r}&0\\0&e^{2r}.
\end{psmallmatrix}}
When $q=0$, we obtain the squeezed state, which is pure. We will write the pure squeezed state as
\begin{equation}\label{eq:squeezedvac}
    |\zeta\rangle = \hat S_\zeta|0\rangle.
\end{equation}

Since Gaussian states are fully characterized by their displacement vector and covariance matrix, they form a five-parameter family of states: ${(q_0,p_0, r, \phi, q)}$. One can therefore write the most general single-mode Gaussian state as
\begin{equation}
    \rhoG = \hat D (\alpha) \hat S (\zeta) \rhoh_{\text{Th}} \hatd S (\zeta) \hatd D (\alpha)
\end{equation}

With these elements in hand, we can turn to the derivation of the Wigner function for a Gaussian state transformed by a cubic phase gate.


\section{The Wigner function of cubic Gaussian states}\label{sec:cubic}
We consider the cubic gate,
\begin{equation}\label{eq:cubicgate}
   \hat \Gamma_\gamma=\exp(-i\gamma \hat q^3),\quad \gamma\geq0,
\end{equation}
and its action on a state  $\rhoh$:
\begin{equation}\label{eq:rhogamma}
\rhoh_\gamma=\hat \Gamma_{\gamma} \rhoh \hatd \Gamma_{\gamma}.
\end{equation}
We write $W_{\gamma}$ for the Wigner function of $\hat\rho_\gamma$. We shall concentrate on the case where $\rhoh$ is a Gaussian state {$\rhoG$}, in which case we write:
\begin{equation}\label{eq:rhoGgamma}
\rhoh_{\textrm{G}, \gamma}=\hat \Gamma_{\gamma} \rhoh_G \hatd \Gamma_{\gamma};
\end{equation}
we denote its Wigner function by $W_{\textrm{G}, \gamma}$.  We shall refer to the states $\rhoh_{\textrm{G}, \gamma}$ as \textit{cubic Gaussian states} as they are Gaussian states transformed under the cubic phase gate. Note that the cubic Gaussian states are not Gaussian: the cubic gate, which has a cubic Hamiltonian, ``degaussifies'' the Gaussian states $\rhoG$. 
We will write the pure cubic squeezed states as
\begin{equation}\label{eq:cubicsqueezedvac}
    |\gamma, \zeta\rangle =\Gamma_\gamma S_\zeta|0\rangle.
\end{equation}
The Wigner function of an arbitrary cubic Gaussian state can be computed explicitly using Eq.~\eqref{eq:WignerGaussian},   Eq.~\eqref{eq:kernelsW} and Eq.~\eqref{eq:Winvkernel}.
The result is
\begin{equation}\label{eq:WignerLambda}
\WGgamma(q,p)=\frac1\pi\sqrt{\frac2{\pi\det V}}\exp\Big[-\frac1{2\sigma_{qq}}(q-q_0)^2\Big]\Lambda(x,\lambda),
\end{equation}
where 
\begin{eqnarray}\label{eq:definitionvariablesnew}
 \lambda &=& 6\gamma \left(\frac{\sigma_{qq}}{\det V} \right)^{3/2},\label{eq:lambdadef}\\
x&=& 2\sqrt{\frac{\sigma_{qq}}{\det V}} \left( (p-p_0) + 3 \gamma q^2 - \frac{\sigma_{qp}}{\sigma_{qq}} (q-q_0) \right),\label{eq:xdef}\\
    \Lambda(x,\lambda)&=&\int \exp[ixs-\frac12 s^2+i\lambda\frac{s^3}{3}]\rd s.
\end{eqnarray}
One readily checks that, when $\gamma\to0$, and hence $\lambda\to0$, one recovers the Wigner function of the underlying gaussian state $\rhoG$. (The details of these and the following computations are given in Appendix~\ref{app:wignercubic}.) The parameter $\lambda$ will play a central role in what follows. 

For our purposes, it will be convenient to have an alternative expression of  $W_{\textrm G, \gamma}(q,p)$ in terms of the Airy function:
\begin{equation}\label{eq:wignercubicgaussianstate}
\begin{split}
     \WGgamma(q,p) & = 2 \sqrt{ \frac{2}{ \pi \det{V}}} \exp{ \left[-  \frac{1}{2 \sigma_{qq}} (q-q_0)^2 \right]}  |\lambda|^{-1/3}\exp(-\frac{1}{24\lambda^2})\exp(\frac{\eta}{2\lambda^{2/3}}) \textrm{Ai}(\eta).
     \end{split}
\end{equation}
where
\begin{equation}\label{eq:etadef}
  \eta=\lambda^{-1/3}\left(x+\frac1{4\lambda}\right),
\end{equation}
and Ai is the Airy function \cite{chin_dispersion_nodate,schleich_quantum_2001,brunelli_linear_2019}:
\begin{equation}\label{eq:Airy}
   \textrm{Ai}(x) = \frac{1}{2 \pi} \int \exp{\left(\frac{i y^3}{3}  + ix y \right) } \rd y.
\end{equation}
The above expression is valid for all cubic Gaussian states, and  was previously obtained  in the special case where $\rhoG$ is a displaced squeezed vacuum \cite{ghose_non-gaussian_2007,takagi_convex_2018,brunelli_linear_2019}.

\begin{figure}[h!]
  \subfloat[\label{fig:WignerLambda0212}]{\includegraphics[width=4cm]{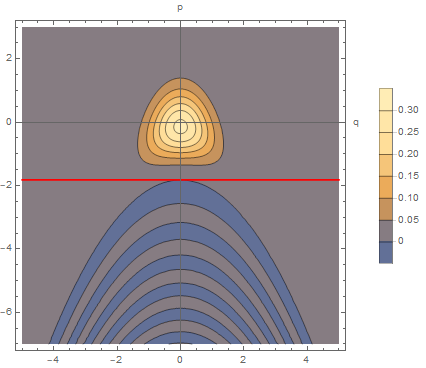}}
  \hspace{0.1cm}
  \subfloat[\label{fig:WignerLambda0075}]{\includegraphics[width=4cm]{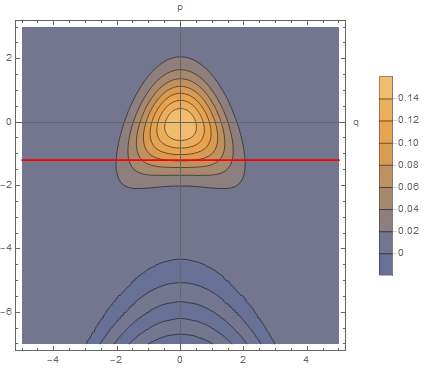}}
  \hspace{0.1cm}
  \subfloat[\label{fig:WignerNegLambda426}]{\includegraphics[width=4cm]{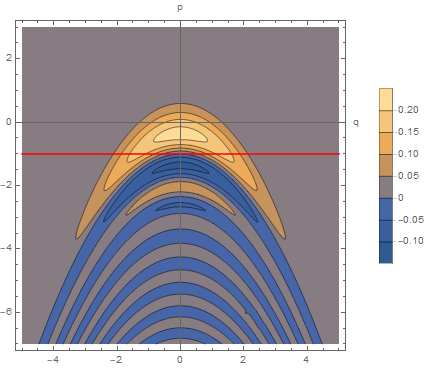}}
     \hspace{0.1cm}
  \subfloat[\label{fig:WignerNegLambda151}]
  {\includegraphics[width=4cm]{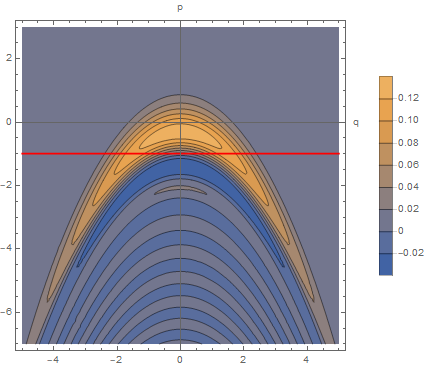}}
   \caption{Wigner function of centered  cubic Gaussian states (Eq.\eqref{eq:wignercubicgaussianstate}) with thermal and squeezing parameters as indicated and $\gamma = 0.1$. 
   Above the solid (red) line, the Wigner function is guaranteed to be positive (see Eq.(\ref{eq:WigPositiveHalfPlane}).
   (a) Cubic vacuum state; $\lambda = 0.212$; $\Ncal_\lambda \approx 1.0046$.
   (b) Cubic thermal state; $q = 1/3$; $\lambda = 0.075$; $\Ncal_\lambda \approx 1.0000$.
   (c) Cubic squeezed state; $\zeta=-1$; $\lambda = 4.26$; $\Ncal_\lambda \approx 1.9309$.
   (d) Cubic thermal squeezed state; $q=1/3$; $\zeta=-1$; $\lambda = 1.51$; $\Ncal_\lambda \approx 1.3571$. 
}
   \label{fig:cubicphaseWignerFct1}
\end{figure}

 The Airy function is known to have a sequence of zeroes
$\dots \eta_m<\eta_{m-1}<\dots<\eta_0<0,$
 and to be positive for all $\eta>\eta_0$. For further reference, we recall that $|\eta_0|\simeq 2.34$. It follows that the Wigner function $W_{G,\gamma}(q,p)$ vanishes identically on  the parabolae
$\eta=\eta_m$ in phase space.
 More precisely, the Wigner function oscillates from positive values between the parabola $\eta=\eta_{2n}$ and $\eta=\eta_{2n-1}$ and negative values between the parabolae $\eta=\eta_{2n+1}$ and $\eta_{2n}$ and it is positive in the entire region $\eta>\eta_0$. This behaviour can be observed in the plots of Fig.~\ref{fig:cubicphaseWignerFct1}.
 
 The above considerations imply that if $W_{\textrm G,\gamma}(q,p)$ is negative, then $\eta<\eta_0$, \textit{i.e.} $(\lambda>0)$
 \begin{equation}
 \begin{split}
      2\sqrt{\frac{\sigma_{qq}}{\det V}}\left((p-p_0)-\frac{\sigma_{qp}}{\sigma_{qq}}(q-q_0)+3\gamma q^2\right) &\leq -\frac1{4\lambda}+\lambda^{1/3}\eta_0<0.
 \end{split}
 \end{equation}
 Consequently, if 
  \begin{equation}\label{eq:WigPositiveHalfPlane}
  \begin{split}
       2\sqrt{\frac{\sigma_{qq}}{\det V}}\left((p-p_0)-\frac{\sigma_{qp}}{\sigma_{qq}}(q-q_0)\right) &> -\frac1{4\lambda}+\lambda^{1/3}\eta_0,
  \end{split}
 \end{equation}
 then $W_{G,\gamma}(q,p)>0$. 
 This last inequality defines a half plane. It is readily checked that  it contains the disc of radius $R_*$ centered on  the point $(q_0, p_0)$, where 
 \begin{equation}
 \begin{split}
       x_0=-\frac1{4\lambda}+\lambda^{1/3}\eta_0<0, & \quad R_*=\frac12|x_0|\left[\frac{1}{\det V}\left(\sigma_{qq}+\frac{{\sigma_{qp}}^2}{\sigma_{qq}}\right) \right]^{-1/2}. \\
 \end{split}
 \end{equation}
It follows that, when $\gamma>0$ tends to zero,  while $\rhoG$ is fixed, $R_*\sim \gamma^{-1}\to+\infty$, so that the region where $W_{\textrm G,\gamma}(q,p)$ is positive gradually invades the entire $(q,p)$-plane: the negativities of the Wigner function are pushed out to infinity.  This is not surprising: indeed, the limiting Gaussian Wigner function is everywhere positive, except at infinity, where it vanishes. A perturbation of this state by a cubic gate with small $\gamma$ can be expected to induce negativities only there where the original Gaussian is small, which is close to infinity. This explains, why, when $\gamma$ is small, there is a large ball of radius $R_*\sim \gamma^{-1}$ around $(q_0, p_0)$ where the Wigner function of the cubic Gaussian state remains positive.
In addition, when $\gamma$ is small, the frequency of oscillations in $W_{\text{G},\gamma}$ increases, since the successive parabolae $\eta=\eta_m$ are separated by a distance of order $\gamma^{1/3}$. But, since the amplitude of these fast oscillations is very small, we will show in the next section, that for small $\gamma$, the Wigner negativity (see Eq.~\eqref{eq:wigneg}) of the cubic Gaussian state is exceedingly small (See Fig.~\ref{fig:NegativityBounds}). 

The location and behaviour of the negative values of the Wigner function of cubic Gaussian states is very different from that of the photon added or substracted Gaussian states \cite{hertz_decoherence_2023}. As an example, under photon addition of squeezed thermal states, the Wigner function develops negative values at the origin of phase space, not at infinity. In addition, at large squeezing, the Wigner negative volume saturates. In contrast, as can be observed in Fig.(\ref{fig:NegativityBounds}), in the case of the cubic squeezed state, the Wigner negativity increases monotonically and continuously with the squeezing and with the cubicity through the parameter $\lambda$.

\begin{figure}
\centering
  \subfloat[\label{fig:WignerDispLambda0212}]{\includegraphics[width=4cm]{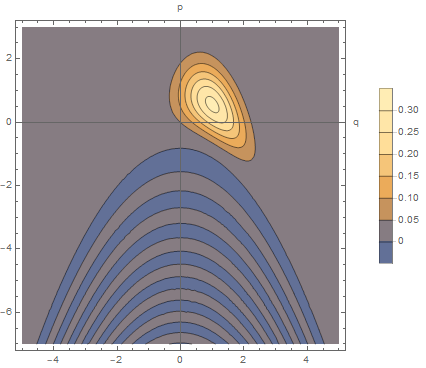}}
  \hspace{0.1cm}
  \subfloat[\label{fig:WignerDispLambda0075}]{\includegraphics[width=4cm]{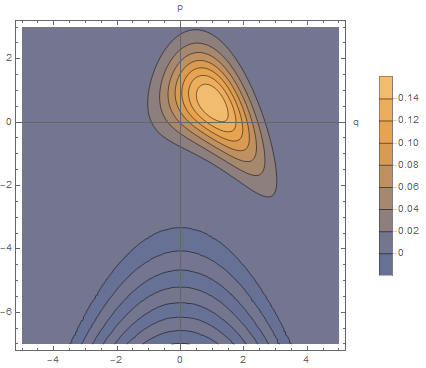}}
  \hspace{0.1cm}
  \subfloat[\label{fig:WignerDispLambda426}]{\includegraphics[width=4cm]{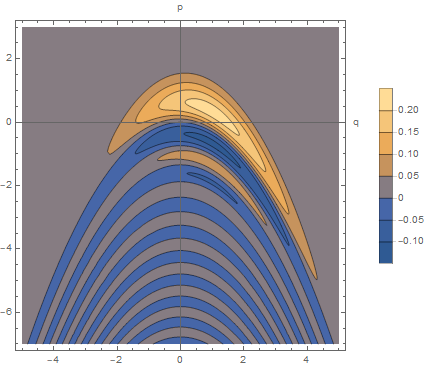}}
     \hspace{0.1cm}
  \subfloat[\label{fig:WignerDispLambda151}]
  {\includegraphics[width=4cm]{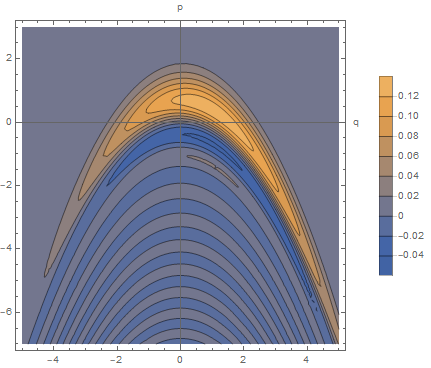}}
   \caption{Wigner function of the  cubic displaced Gaussian state [Eq.(\ref{eq:wignercubicgaussianstate}) ] with displacement $q_0 = p_0 = 1$ and $\gamma = 0.1$. (a) Cubic displaced vacuum state; $\lambda = 0.212$; $\Ncal_\lambda \approx 1.0046$. (b) Cubic displaced thermal state; $q = 1/3$; $\lambda = 0.075$; $\Ncal_\lambda \approx 1.0000$. (c) Cubic displaced squeezed state; $\zeta=-1$; $\lambda = 4.26$; $\Ncal_\lambda \approx 1.9309$. (d) Cubic displaced squeezed thermal state; $q=1/3$; $\zeta=-1$; $\lambda = 1.51$; $\Ncal_\lambda \approx 1.3571$.
}\label{fig:cubicphaseWignerFct2}
\end{figure}

In Fig.~\ref{fig:cubicphaseWignerFct2} we show the Wigner function of the cubic Gaussian states with the same covariance matrix as those shown in Fig.~\ref{fig:cubicphaseWignerFct1}, but displaced by $q_0=1, p_0=1$. Comparing the figures, one observes that the Wigner functions are not simply displaced versions of the corresponding Wigner functions with centered Gaussians, but that the displacement followed by a cubic gate also deforms the resulting Wigner function. The observed shift in the positive part of the Wigner function, and hence of its barycenter, can be qualitatively understood by observing that
\begin{equation}\label{eq:meandisplacement}
   \tr(\hat{q}\,\rhoh_{\text G,\gamma})=q_0,\quad   \tr ( \hat{p}\, \rhoh_{\text G,\gamma}) = \mean{\hat{p}}_{\text G} - 3 \gamma \mean{\hat{q}^2}_{\text G}= p_0-3\gamma (\sigma_{qq}+q_0^2).
\end{equation}

\section{The Wigner Negativity of the cubic Gaussian states}\label{sec:wignerneg}

As explained in the introduction, negativity of the Wigner function of a state $\rhoh$ is associated with nonclassical features of the state.  To measure the amount of Wigner negativity, a much used figure of merit is the so-called Wigner negativity of the state \cite{kenfack_negativity_2004}, that we define as
\begin{equation}\label{eq:wigneg}
    \mathcal N (\rhoh)={\frac12}\iint |W|(q,p)\rd q\rd p.
\end{equation}
Noting that one can write the Wigner function as 
\begin{equation}
W(q,p)=W_+(q,p)-W_-(q,p),
\end{equation}
where $W_+$ and $W_-$ designate its positive and negative parts,
one observes that

\begin{equation}\label{def:wignerneg_positivenegative}
\begin{split}
    \mathcal{N}(\rhoh) 
        & = 1 + \iint W_-(q,p) dq dp
         = \iint W_+(q,p) dq dp  - 1.
\end{split}
\end{equation}
One often refers to $\iint W_-(q,p)\rd q\rd p$ as the Wigner negative volume \cite{kenfack_negativity_2004}, which is, in view of what precedes, simply related to the Wigner negativity.

\begin{figure}[t!]
  \subfloat[\label{fig:NegFullLogLog}]{\includegraphics[width=5.5cm]{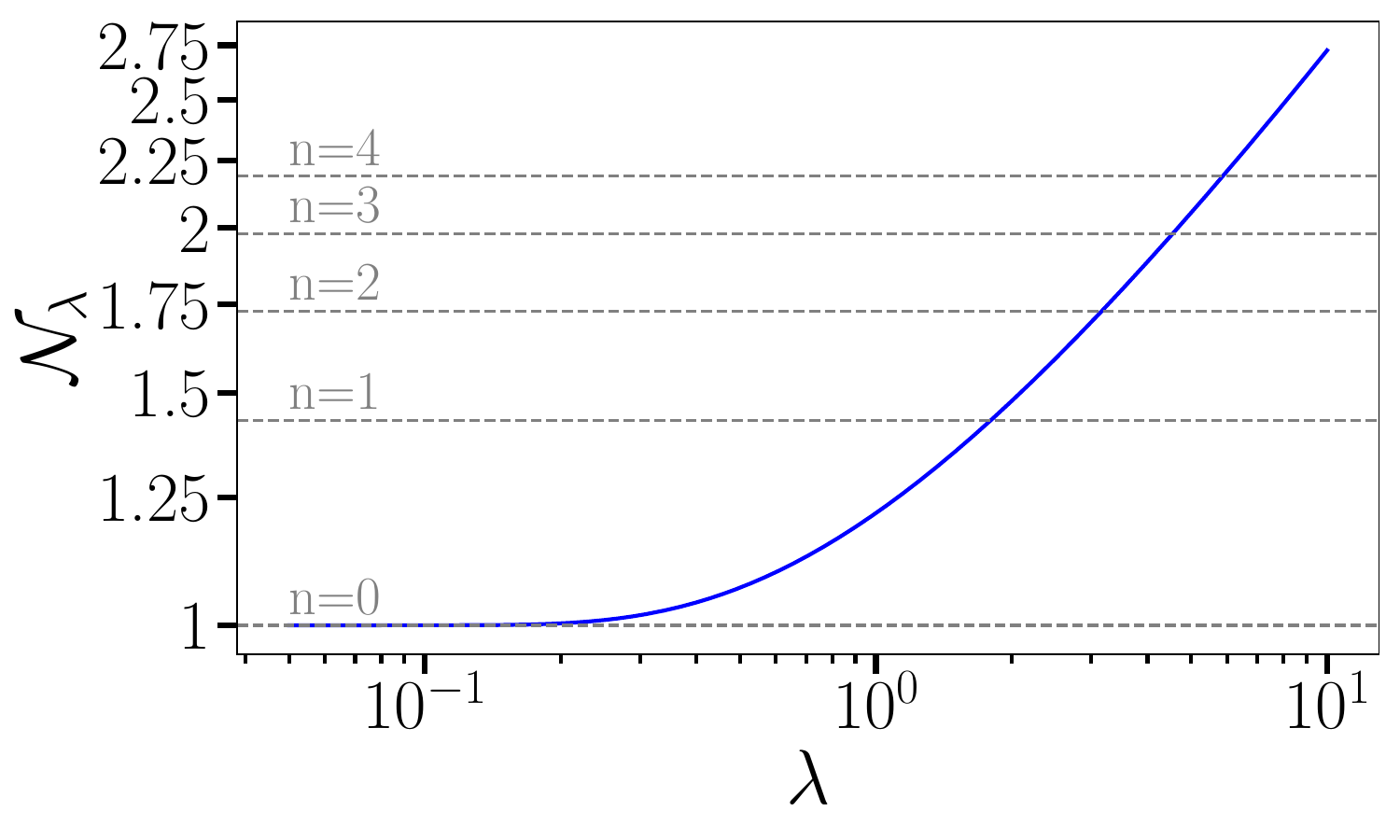}}
  \hspace{0.1cm}
  \subfloat[\label{fig:NegSmallLog}]{\includegraphics[width=5.5
  cm]{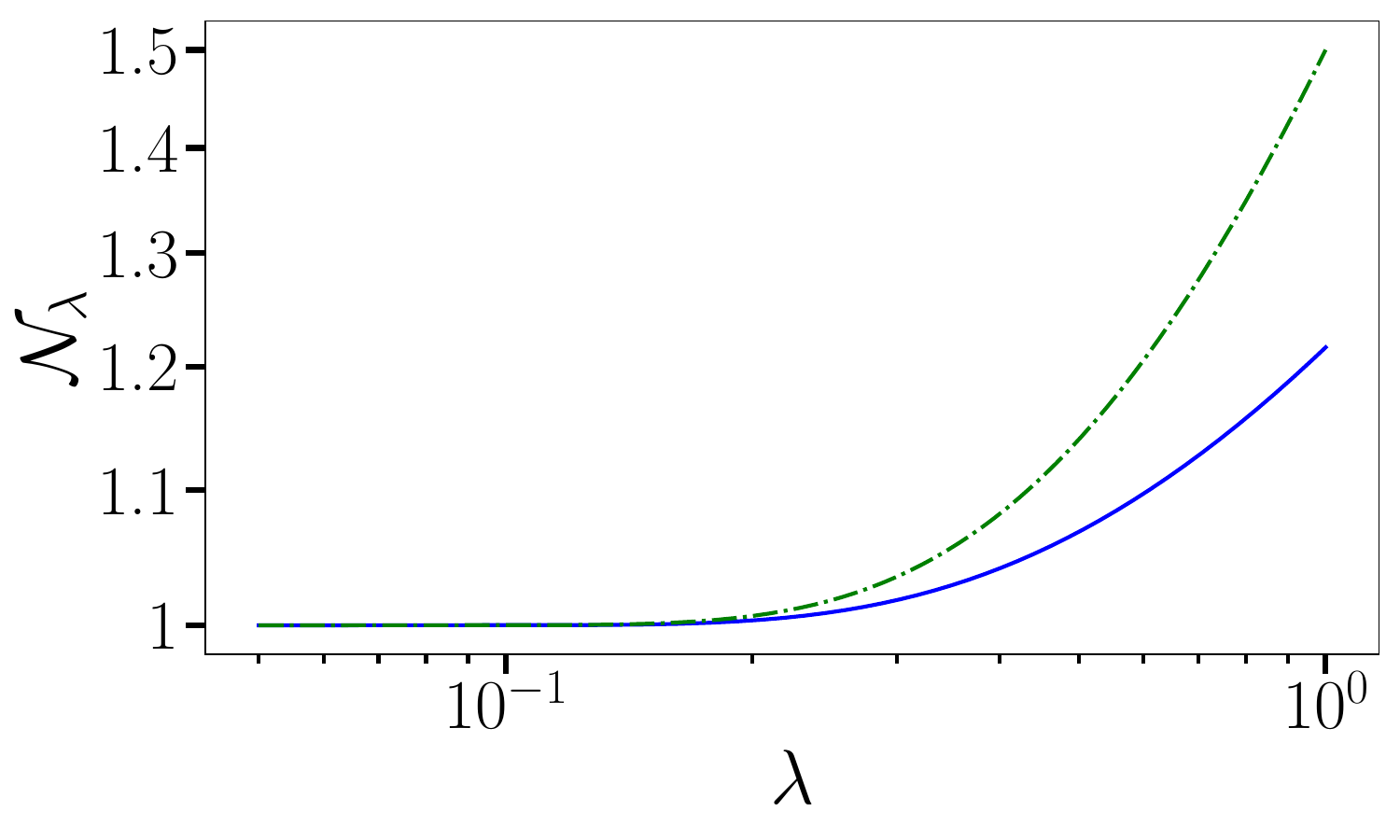}}
  \hspace{0.1cm}
  \subfloat[\label{fig:NegLargeLog}]{\includegraphics[width=5.5cm]{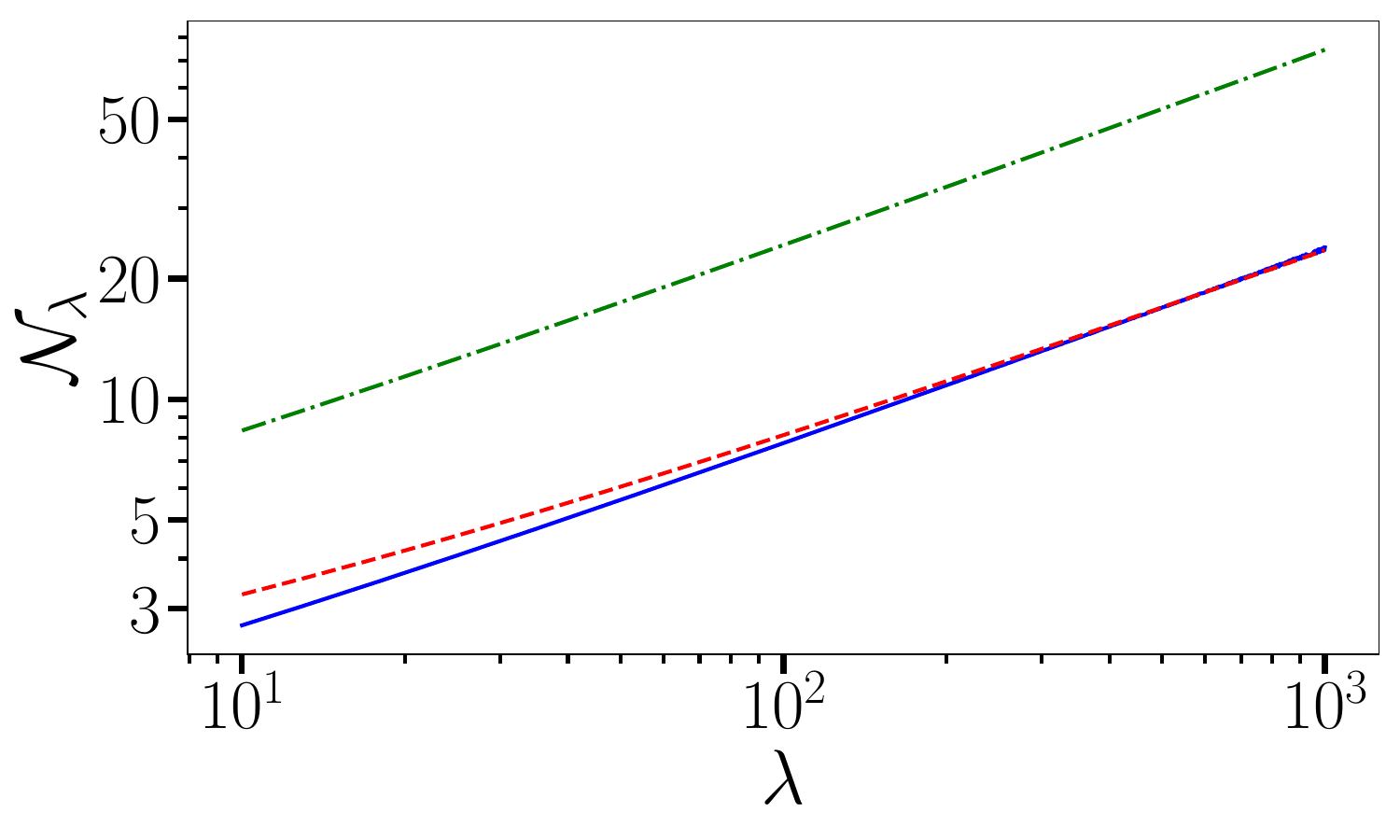}}
   \caption{ Negativity of cubic Gaussian states  as a function of $\lambda$ from Eq.(\ref{eq:wignerneg}) in solid blue line. 
   (a) Log-Log plot for $\lambda \in \left[ 0.05, 10  \right]$. The dotted lines indicate the negativity of the Fock states $0$ to $4$ (see Table. \ref{table:lambdan})
   (b) Log-linear plot for small values of $\lambda \in \left[ 0.05, 1 \right]$. The dot-dashed green line is the upper-bound from Eq. (\ref{eq:wignernegupperboundsmallgamma}).
   (c) Log-log plot for large values of $\lambda \in \left[ 10, 1000 \right]$. The upper dot-dashed green line is the upper-bound from Eq.(\ref{eq:wignernegupperboundlargegamma}) and the dashed red line is the asymptotic approximation $0.75 \sqrt{\lambda}$.}
   \label{fig:NegativityBounds}
\end{figure}

In this section, we shall study the Wigner negativity of general cubic Gaussian states. Using Eq.~(\ref{eq:wignercubicgaussianstate}) and performing a change of variables in the integral (see Appendix~\ref{app:bounds} for details), it can be easily expressed in the form 
 \begin{equation}\label{eq:wignerneg}
    \begin{split}
        \mathcal N(\rhoGgamma) & = \exp\left(-\frac{1}{24\lambda^2}\right) \int \exp\left(\frac{\eta}{2\lambda^{2/3}}\right) |\textrm{Ai}(\eta)| \rd \eta=:\mathcal N_\lambda.
    \end{split}
\end{equation}
We are not aware of an analytical expression for the remaining integral because of the presence of the absolute value of the Airy function. The integral can of course be evaluated numerically but both for large and small $\lambda$ an accurate evaluation is complicated by the oscillations in the Airy function. Plots of $\mathcal N_\lambda$ on different scales are provided in Fig. \ref{fig:NegativityBounds}.

One observes that the Wigner negativity Eq.(\ref{eq:wignerneg}) of the six-parameter family of cubic Gaussian states depends on one single parameter $\lambda$, defined in Eq.~\eqref{eq:lambdadef}. In particular, whereas the Wigner function of $\rhoGgamma$ itself does depend nontrivially on the displacement vector $(q_0, p_0)$ of the Gaussian state $\rhoG$, the Wigner negativity of the cubic Gaussian state $\rhoGgamma$  does not. This is in fact  a general feature, as we show in Appendix \ref{app:wignernegdisplinvariance}: the Wigner negativity of a cubic state $\Gamma_\gamma\hat\rho\Gamma^\dagger_{\gamma}$  is identical to that of the cubic state  $\Gamma_\gamma D(\alpha)\hat\rho D(\alpha)^\dagger\Gamma^\dagger_{\gamma}$, for any displacement $\alpha$.

In absence of an explicit analytical expression for the Wigner negativity of cubic Gaussian states,  it is instructive to determine simple asymptotic estimates for this quantity, both at large and small values of $\lambda$. We derive such expressions in Appendix~\ref{app:bounds} and give the results here.  
First, one readily obtains,  using the fact that the Airy function is bounded (see Appendix~\ref{app:bounds} for details), the following  upper bound on the Wigner negativity of the cubic Gaussian state: 
\begin{equation}\label{eq:wignernegupperboundsmallgamma}
\mathcal N_\lambda\leq 1+2M_0 2\lambda^{2/3}\exp\left(-\frac1{24\lambda^2}\right)\exp\left(-\frac{|\eta_0|}{2\lambda^{2/3}}\right),
\end{equation}
where $M_0=\max_{\eta<\eta_0} |\textrm{Ai}(\eta)| \approx 0.42$; recall that $\eta_0$ is the largest (negative) eigenvalue of the Airy function. 
Note that while this estimate holds for all values of $\lambda$, it is of interest mainly for relatively small values of $\lambda$, where it shows that the Wigner negative volume is extremely small. For example, for values of $\lambda \leq 0.3$, the negativity $(\mathcal N_\lambda -1)$ created by the cubic gate is less than 5\%  of the negativity $(\mathcal N(|n\rangle\langle n|)-1)$ of the first Fock state $|1\rangle$.

For large $\lambda$, the bound Eq.~\eqref{eq:wignernegupperboundsmallgamma} shows that $\mathcal N_\lambda$ grows at a rate lower than $\lambda^{2/3}$. However, a better bound can be derived (see Appendix~\ref{app:bounds} for details) for large values of $\lambda$ by making use of the asymptotic expansion of the Airy function:
\begin{equation}\label{eq:wignernegupperboundlargegamma}
\begin{split}
      \mathcal N_\lambda & \leq 1+2\frac{2^{3/4}}{\sqrt\pi} \Gamma\left(\frac{3}{4}\right) \lambda^{1/2}\exp\left(-\frac1{24\lambda^2}\right).
\end{split}
\end{equation}
This shows that the growth rate of $\mathcal N_\lambda$ is at most $\lambda^{1/2}$. The two upper bounds Eqs.(\ref{eq:wignernegupperboundsmallgamma}-\ref{eq:wignernegupperboundlargegamma}) are displayed in Fig. \ref{fig:NegativityBounds}. More precisely, the numerics further indicate that the growth of $\mathcal N_\lambda$ for large $\lambda$ is of the order $\lambda^{1/2}$, a result that can probably be proven rigourously by exploiting the finer properties of the Airy function\footnote{We thank O. Lafitte for pointing this out to us.}. We find numerically that
\begin{equation}\label{eq:Nlambdaasympt}
    \mathcal N_\lambda=1+c \sqrt \lambda,
\end{equation}
where $c\approx 0.745$.

\section{The quadrature coherence scale of the cubic Gaussian states}\label{sec:qcs}

The quadrature coherence scale $\Ccal(\rhoh)$ (QCS) of a state $\rhoh$, introduced in~\cite{de_bievre_measuring_2019, hertz_quadrature_2020}, is defined as
\eq{\Ccal^2(\rhoh)=\frac1{2\mathcal{P}(\rhoh)}
\left(\tr[\rhoh,\hat q][\hat q,\rhoh]+\tr[\rhoh,\hat p][\hat p,\rhoh]\right)\label{eq:QCSdef}, }
where   $\mathcal{P}(\rhoh)=\tr\rhoh^2$ is the purity of the state $\rhoh$. It can be seen to be a measure for how far from the diagonal the $\hat q$- and $\hat p$-coherences of $\rho$ live~\cite{hertz_quadrature_2020}. For our purposes here, its main property, shown in~\cite{hertz_quadrature_2020}, is that it provides a quantitative measure for the decoherence rate of the state $\rhoh$, when it is coupled to a thermal bath through a Lindblad equation. More precisely, if $\tau$ denotes the decoherence time, one finds that 
\begin{equation}\label{eq:tau_QCS}\tau\simeq \Ccal^{-2}.
\end{equation}
In other words, states with a large QCS decohere quickly. For further properties of the QCS, and in particular the fact that it provides a witness of nonpositivity of the Glauber-Sudarshan $P$-function, as well as its behaviour on numerous benchmark states, we refer to
~\cite{de_bievre_measuring_2019, horoshko_thermal-difference_2019, hertz_quadrature_2020, hertz_relating_2020, hertz_decoherence_2023, hertz_quadrature_2024}. These works show that the QCS, contrary to many other measures or witnesses of $P$-nonpositivity, is often easy to compute; some examples are given below. Let us in addition point out that an interferometric protocol for direct measurement of the QCS has been proposed in~\cite{griffet_interferometric_2023} and experimentally tested in~\cite{goldberg_measuring_2023}. 

For pure states, a simple computation starting from \eqref{eq:QCSdef} shows that 
\begin{eqnarray}\label{eq:pureQCS}
\Ccal^2(\rhoh)& = (\Delta \hat q)^2 +(\Delta \hat p)^2 &=2\langle (a^\dagger - \langle a^\dagger \rangle) ( a - \langle a \rangle) \rangle+1, \\
& & = 2 (\mean{a^\dagger a} - \mean{a^\dagger}\mean{a}) + 1.
\end{eqnarray}
 which is the so-called total noise of $\hat\rho$. 
For a single-mode Gaussian state $\rhoG$, pure or mixed, one finds~\cite{hertz_quadrature_2020, hertz_relating_2020}
 \begin{equation}\label{eq:QCSGaussian}
 \QCStwoG=\Ccal^2(\rhoG)=\frac1{2}\Tr V^{-1}.
 \end{equation}
 For example, the 
 squeezed thermal states have $\Ccal^2_{\textrm{SqTh}}=\frac{1-q}{1+q}\cosh r$. Note the fast growth of the QCS with the squeezing parameter $r$: strongly squeezed Gaussian states are very sensitive to environmental decoherence.

In order to assess the sensitivity of the cubic Gaussian states to decoherence, we evaluate and analyze their QCS in this section. Using that, for any unitary operator $\hat U$ and any self-adjoint operator $\hat A$, one has that 
\begin{equation}
    [\hat U \rhoh \hat U^\dagger, \hat A]= \hat U[\rhoh,\hat U^\dagger \hat A \hat U]\hat U^\dagger,
\end{equation}
and
\begin{equation}\label{eq:momentumgamma}
    \hatd \Gamma_{\gamma}\hat p \hat \Gamma_\gamma=\hat p-3\gamma \hat q^2,
\end{equation}
one finds that 
\begin{eqnarray}\label{eq:QCScubicgaussian}
    \mathcal C^2_{\textrm G, \gamma}&=&\frac{1}{2  \mathcal{P}(\rhoG)}\left(\Tr([\rhoG,\hat q][\hat q,\rhoG]+[\rhoG, \hat p-3\gamma\hat q^2][\hat p-3\gamma \hat q^2,\rhoG])\right), \\
     &=&\mathcal C^2(\rhoG)- 3\gamma  \frac{\tr[\rhoG,\hat p][\hat q^2,\rhoG]}{\mathcal{P}(\rhoG)} +\frac92 \gamma^2 \frac{\tr[\rhoG,\hat q^2][\hat q^2,\rhoG]}{ \mathcal{P}(\rhoG)} ,
\end{eqnarray}
where $\mathcal{P}(\rhoG) = 1/\sqrt{\det V}$. 
The linear term in $\gamma$ can be simplified as follows. Recall that, for $z_0=\frac1{\sqrt2}(q_0+ip_0)$, one has 
\begin{equation}
   \hatd D(z_0)\hat a \hat D(z_0)=\hat a +z_0,\quad \hatd D(z_0)\hat q \hat D(z_0)=\hat q+q_0,\quad \hatd D(z_0) \hat p \hat D(z_0)=\hat p+p_0.
\end{equation}
Now choose $q_0=\langle \hat q\rangle_{\textrm G}$ and $p_0=\langle \hat p\rangle_{\textrm G}$ and define
\begin{equation}
    \hat\rho_{G^0}= \hat D(-z_0)\rhoG \hat D^\dagger(-z_0) = \hatd D(z_0)\rhoG \hat D(z_0),
\end{equation}
which is the centered Gaussian with the same covariance matrix as $\rhoG$. Then
\begin{eqnarray}
 \tr[\rhoG,\hat p][\hat q^2,\rhoG]&=&
 \tr[\hat\rho_{G^0},D^\dagger(z_0)\hat p \hat D(z_0)][ \hat D^\dagger(z_0)\hat q^2 \hat D(z_0),\hat\rho_{G^0}],\\
 &=&
  \tr[\hat\rho_{G^0},\hat p ][(\hat q+q_0)^2,\hat\rho_{G^0}],\\
  &=&\tr[\hat\rho_{G^0},\hat p ][\hat q^2,\hat\rho_{G^0}]+2q_0\tr[\hat\rho_{G^0},\hat p ][\hat q,\hat\rho_{G^0}].
\end{eqnarray}
Since $\hat\rho_{G^0}$ is parity invariant, the first term is easily seen to vanish. Reintroducing $\rhoG$, one then finds that
\begin{equation}
    \tr[\rhoG,\hat p][\hat q^2,\rhoG]=2q_0\tr[\rhoG,\hat p ][\hat q,\rhoG].
\end{equation}
Similarily, we can compute the term quadratic in $\gamma$ as
\begin{eqnarray*}
    \tr[\rhoG,\hat q^2][\hat q^2,\rhoG]&=&
 \tr[\hat\rho_{G^0},D^\dagger(z_0)\hat q^2 \hat D(z_0)][ \hat D^\dagger(z_0)\hat q^2 \hat D(z_0),\hat\rho_{G^0}],\\
 &=&
  \tr[\hat\rho_{G^0},(\hat q+q_0)^2][(\hat q+q_0)^2,\hat\rho_{G^0}],\\
  &=&\tr[\hat\rho_{G^0},\hat q^2 ][\hat q^2,\hat\rho_{G^0}]+4q_0^2\tr[\hat\rho_{G^0},\hat q ][\hat q,\hat\rho_{G^0}].
\end{eqnarray*}
We conclude that
\begin{equation}\label{eq:QCScubicdispl}
   \mathcal C^2_{\textrm G, \gamma}=\mathcal C^2(\hat\rho_{G^0}) +\frac92\gamma^2 \left( \frac{\tr[\hat\rho_{G^0},\hat q^2][\hat q^2,\hat\rho_{G^0}]}{\mathcal{P}(\rhoG)} + 4q_0^2 \frac{\tr[\hat\rho_{G^0},\hat q ][\hat q,\hat\rho_{G^0}]}{\mathcal{P}(\rhoG)} \right)-6\gamma q_0 \frac{\tr[\hat\rho_{G^0},\hat p][\hat q,\hat\rho_{G^0}]}{\mathcal{P}(\rhoG)},
\end{equation}
where we have explicitly shown the effect of the displacement on the QCS of the cubic Gaussian state $\rho_{G,\gamma}$. One can notice that this derivation only uses the assumption that the centered state is parity invariant. Hence, the resulting expression   is valid for all parity invariant centered states, not only for Gaussian ones. 

The QCS of a cubic Gaussian state is computed in Appendix~\ref{app:QCScubic} and found to be 
\begin{eqnarray}\label{eq:QCScubiccovelements}
    \mathcal C^2_{\textrm G, \gamma}&=&\mathcal C^2(\rhoG) + 18 \gamma^2 \left( \frac{\sigma_{qq}^2}{\det V} + 2 q_0^{2} \frac{\sigma_{qq}}{\det V} \right) - 6 \gamma q_0 \frac{(- 2 \sigma_{qp})}{\det V}, \\
    &=& \frac{1}{\det V} \left(\sigma_{pp} + \sigma_{qq} + 18 \gamma^2\sigma_{qq}^2 + 12 q_0  \gamma \left(3 \gamma q_0 \sigma_{qq} + \sigma_{qp}\right) \right).
\end{eqnarray}
In the special case of centered Gaussian states, the QCS of the cubic centered $(q_0 = 0)$ Gaussian state reduces to  :
\begin{equation}\label{eq:QCSGaussianGamma}
\begin{split}
     \mathcal C^2_{\textrm G^0, \gamma}  =& \left(\frac{1-q}{1+q}\right) \cosh(2r) + \frac92 \gamma^2 (\cosh(2r) - \cos(\phi) \sinh(2r))^2.
\end{split} 
\end{equation}
In contrast to the Wigner negativity $\mathcal{N}_\lambda$ Eq.(\ref{eq:wignerneg}), the expression for the QCS is simple to obtain and provides a completely explicit dependence on the model parameters. In particular, the quadratic dependence on $\gamma$ of the QCS can be easily read from the above equation without the need for an asymptotic analysis as for the Wigner negativity. 



\section{Minimizing the decoherence time: how expensive is a large Wigner negativity?}\label{s:optimalCGstates}
We now consider the following question. Suppose we wish to prepare a cubic Gaussian state with a given negativity: which one of those states has the smallest QCS, and hence the longest decoherence time? We call these states the optimal cubic Gaussian states and study their properties here.

We start by observing that, whereas $\mathcal N(\rhoGgamma)$ depends only on $\lambda$, $\mathcal C^2(\rhoGgamma)$ has a more complex (but completely explicit) dependence on the six parameters of the family of cubic Gaussian states, as is clear from Eq.~\eqref{eq:QCSGaussianGamma}. For the purposes of the optimisation problem at hand, we eliminate $\gamma$ from this expression, in favour of $\lambda$, using the relation Eq.~\eqref{eq:definitionvariablesnew}. One finds
\begin{equation}\label{eq:QCSGaussianCentered}
\begin{split}
    \mathcal C^2_{\textrm G^0,\lambda} & = \left(\frac{1-q}{1+q}\right) \cosh(2r) +  \lambda^2 \left(\frac{1+q}{1-q}\right)^3 \frac{1}{\cosh(2r) - \cos(\phi) \sinh(2r)}.
\end{split}
\end{equation}
and
\begin{equation}\label{eq:QCSGaussianLambda}
\begin{split}
    \mathcal C^2_{\textrm G, \lambda} = &  \mathcal C^2_{\textrm G^0,\lambda}  + 2 \lambda q_0 \left( \left(\frac{\lambda q_0}{2} \right) \left(\frac{\det V}{\sigma_{qq}}\right)^2 + \left(\frac{\det V}{\sigma_{qq}}\right)^{1/2} \frac{\sigma_{qp}}{\sigma_{qq}} \right), \\   
         = &  \mathcal C^2_{\textrm G^0,\lambda}  + 4 \lambda q_0  \left( \lambda q_0 \left(\frac{1+q}{1-q}\right)^{2} \frac{1}{(\cosh(2r) - \cos(\phi) \sinh(2r))^{2}} + \frac{1}{\sqrt{2}} \left(\frac{1+q}{1-q}\right)^{1/2}\frac{\sinh(2r) \sin(\phi)}{(\cosh(2r) - \cos(\phi) \sinh(2r))^{3/2}} \right).
\end{split}
\end{equation}
The  cubic Gaussian states with the same value  of $\lambda$, given by
\begin{equation}\label{eq:lambdagamma}
\begin{split}
  \lambda& = \frac{3}{\sqrt{2}}\left(\frac{1-q}{1+q}\right)^{3/2}\gamma \exp(3r),
  \end{split}
\end{equation}
all have the same Wigner negativity $\mathcal N_\lambda$. In other words, to produce a cubic Gaussian state with a given value of Wigner negativity, one can adjust $\gamma$, for a given Gaussian state, or if one is experimentally limited to small values of $\gamma$, one can try to adjust the $\hat q$-variance $\sigma_{qq}$ of the underlying Gaussian state $\hat\rho_{\textrm G^0}$ through squeezing, as has been suggested in the literature~\cite{zheng_gaussian_2021, kala_cubic_2022,eriksson_universal_2023}.  However, as is clear from the first term in the right hand side of Eq.~\eqref{eq:QCSGaussianCentered}, large squeezing leads to a large value of the QCS, and hence to a short decoherence time for the state.  It is then natural to ask which cubic Gaussian centered state yields the lowest QCS, for a given value of $\lambda$, and hence of the Wigner negativity $\mathcal N(\hat\rho_{\textrm G^0, \lambda})=\mathcal N_\lambda$. To answer this question, we need to minimize $\mathcal C^2_{\textrm G^0,\lambda}$ over the parameters $r,q,\phi$ determining $\hat\rho_{\textrm G^0}$.  We write
\begin{equation}
\mathcal C^2_{\opt, \lambda}=\min_{q,r,\phi} \mathcal C^2_{\textrm G^0,\lambda}.
\end{equation}
We now determine the values $r_{\opt}, q_{\opt}, \phi_{\opt}$ of the centered Gaussian state $\hat\rho_{\textrm G^0}$ at which this minimum is reached. Inspection of Eq.~\eqref{eq:QCSGaussianCentered} reveals that $\phi_{\opt}=\pi$, which corresponds to cubic $\hat p$-squeezed thermal states ($\zeta=-r$). So we need to determine  
\begin{equation}\label{eq:QCSmin}
\mathcal C^2_{\opt, \lambda}=\min_{q,r} \left[\left(\frac{1-q}{1+q}\right) \cosh(2r) +  \lambda^2 \left(\frac{1+q}{1-q}\right)^3 e^{- 2r}\right].
\end{equation}
A straightforward analysis of the expression in square brackets shows that 
\begin{eqnarray}\label{eq:QCSmin2}
    \mathcal C^2_{\opt, \lambda}&=&(1+2\lambda^2)^{1/2}, \quad \textrm{if}\ \lambda\geq \frac1{\sqrt 2}\label{eq:QCSminlambda},\\
    &=& \sqrt 2,\quad \textrm{if}\ \lambda \leq \frac1{\sqrt2}.
\end{eqnarray}
This implies, for $\lambda\geq1/\sqrt2$, that
\begin{equation}\label{eq:QCSminbound}
    \sqrt2\lambda\leq \mathcal C^2_{\opt, \lambda}< \sqrt2\lambda+1,
\end{equation}
so that, for large $\lambda$, $C^2_{\lambda,\opt}$ behaves like $\sqrt2\lambda$.
The corresponding values of $q_{\opt}$ and $r_{\opt}$ are 
\begin{eqnarray}\label{eq:squeezingmin}
r_{\opt}&=&\frac14\ln(1+2\lambda^2),\ q_{\opt}=0\quad \textrm{if}\ \lambda \geq\frac1{\sqrt2},\\
r_{\opt}&=&\frac14\ln2,\ q_{\opt}=\frac{1-(2\lambda^2)^{1/4}}{1+(2\lambda^2)^{1/4}}\quad \textrm{if}\ \lambda \leq\frac1{\sqrt2}.
\end{eqnarray}

We will refer to a cubic Gaussian state with the parameters $\lambda, r_{\opt}, q_{\opt}$ and $\phi_{\opt}=\pi$ as an optimal cubic Gaussian state and denote it by 
\begin{equation}
  \hat\rho_{\opt, \lambda}=\Gamma_{\gamma_{\opt}}\rhoG \Gamma_{\opt}^\dagger,  
\end{equation}
where $\rhoG$ is the Gaussian state with parameters $r=r_{\opt}, q=q_{\opt}, \phi_{\opt}=\pi$. 
Depending on context, we will also write $\hat\rho_{\opt, \mathcal N}=\hat\rho_{\opt, \lambda}$ with $\mathcal N=\mathcal N_\lambda$. 
For $\lambda\geq \frac1{\sqrt2}=0.71$ the underlying optimal Gaussian state is a squeezed vacuum. Its squeezing parameter $r_{\opt}$ grows only logarithmically with $\lambda$. Expressed in dB, it is
\begin{equation}\label{eq:deltamin}
\delta_{\opt}=\frac{20 ~ r_{\opt}}{\ln10}\approx\frac{20}{2.3}r_{\opt}\approx 8.7\,r_{\opt}\approx\frac{8.7}{4}\ln(1+2\lambda^2).
\end{equation}
The regime where $\lambda\leq 1/\sqrt2$ is of limited interest since for these values, the negativity is very small, as pointed out above. 
Note further that, since $\phi_{\opt}=\pi$, the term in Eq.~\eqref{eq:QCSGaussianLambda} that is linear in the displacement $q_0$ vanishes for $\phi_{\opt}$, so that one sees that a nonvanishing $q_0$ would lead to a value of the QCS larger than $\mathcal C^2_{\lambda, \opt}$.

It is instructive to determine the cubicity of $\rhoh_{\lambda, \opt}$. One finds, from Eq.~\eqref{eq:lambdagamma} that 
\begin{eqnarray}\label{eq:gammamin}
    \gamma_{\opt}&=&\left(\frac{2}{\exp(2r_{\opt})}\right)^{3/2}\frac{\lambda}{6}= \frac{\sqrt2}{3}\frac{\lambda}{(1+2\lambda^2)^{3/4}}, \quad\textrm{if}\ \lambda\geq \frac1{\sqrt2},\\
     \gamma_{\opt}&=&\left(2\frac{1+q_{\opt}}{1-q_{\opt}}\exp(-2r_{\opt})\right)^{3/2}\frac{\lambda}{6}
     = \frac{1}{2^{5/8}}\frac{\lambda^{1/4}}{3}, \quad\textrm{if}\ \lambda\leq \frac1{\sqrt2}.
\end{eqnarray}

We have seen in Eq.\eqref{eq:Nlambdaasympt} above that $\mathcal N_\lambda\simeq 1+ c\lambda^{1/2}$ where $c=0.745\pm 0.005$ for large $\lambda$. 
Hence both $r_{\opt}$ and $\delta_{\opt}$ grow only logarithmically with $\mathcal N_\lambda$, whereas the minimal cubicity $\gamma_{\opt}$ actually decreases with $\lambda$.
Specifically, for large $\lambda$, one finds that
\begin{equation}\label{eq:minNlambda}
 \delta_{\opt}\simeq \frac{8.7}4\ln(1+2 (\frac{\mathcal N_\lambda-1}{c})^4),\quad   \gamma_{\opt}\simeq 2^{-1/4}\frac{1}{3}\lambda^{-1/2}\simeq 2^{-1/4}\frac{1}{3} c\mathcal N_\lambda^{-1}.
\end{equation}
It follows that, whereas the cubicity is responsible for the Wigner negativity, it can be taken very small, provided one is capable of preparing sufficiently strongly squeezed Gaussian states. 
Nevertheless, one needs to note that, in view of Eq.~\eqref{eq:Nlambdaasympt}, the QCS of these optimal cubic Gaussian states grows as   
\begin{equation}\label{eq:QCSminvsNegativity}
 \mathcal C^2_{\opt, \lambda}\simeq \sqrt2\lambda\simeq \frac{\sqrt2}{c^2} \mathcal N_\lambda^2.
\end{equation}
Consequently, for any cubic Gaussian state, one has that
\begin{equation}\label{eq:QCScubicGaussianvsNegativity}
 \mathcal C^2(\rhoh_{\textrm{G}_0,\gamma})\geq  \frac{\sqrt2}{c^2} \mathcal N(\rhoh_{\textrm{G}_0,\gamma})^2.
 \end{equation}
  We have therefore shown that the optimal cubic Gaussian states we introduced provide an interesting tradeoff between Wigner negativity and decoherence time: a large negativity necessarily implies a short coherence time.
This means that, whereas indeed, as has been previously suggested~~\cite{kala_cubic_2022,eriksson_universal_2023,zheng_gaussian_2021, kala:25}, to obtain a given negativity, a small cubicity $\gamma$ can be compensated by a large squeezing, or vice versa, this tradeoff comes at a cost: to obtain a large Wigner negativity, one invariably has a large value of the QCS, hence a short decoherence time for the state. The optimal parameter choices are $\gamma_{\opt}, r_{\opt}$; for large negativity $\mathcal N$, the optimal cubicity $\gamma_{\opt}$ decreases as $\mathcal N^{-1}$  and the optimal squeezing $r_{\opt}$ increases logarithmically in $\mathcal N$ (see Eq.~\eqref{eq:minNlambda}). 



\section{Comparing the optimal cubic Gaussian states to their Fock counterparts}\label{sec:comparingtoFock}

As recalled in the introduction, it has been shown that negativity of the Wigner function  is a necessary condition for quantum advantage and that Wigner negativity is a resource for quantum computation~~\cite{mari_positive_2012, veitch_efficient_2013, pashayan_estimating_2015, albarelli_resource_2018}. This suggests that states with a higher Wigner negativity can provide more quantum advantage. There is however no ``unit'' of Wigner negativity and, since it is an unbounded quantity, no maximal negativity.   In addition, states with the same Wigner negativity can be  different from one another in other respects. With this in mind, we consider in what follows for each $n$ the optimal cubic Gaussian state $\rhoh_{\textrm{opt},\lambda_n}$ that has the same Wigner negativity  as the Fock state $|n\rangle$, that we think of as a benchmark.  
For that purpose, we define $\lambda_n$ through the equation
\begin{equation}\label{eq:lambdan}
    \mathcal N_{\lambda_n}=\mathcal N(|n\rangle\langle n|)=\mathcal N(|n\rangle),
\end{equation} where  $\mathcal N_{\lambda}$ is defined in Eq.(\ref{eq:wignerneg}).
\begin{table}
\begin{tabular}{|c|c|c|c|c|c|c||c|c|c|c|c|c|c|}
\hline
$n$ & $\mathcal{N}(|n\rangle)$ & $\Ccal^2(|n\rangle)$ & $\lambda_n$ & $\Ccal_{\mathrm{opt}, \lambda_n}^2$  & $\delta_{\mathrm{opt},n}$ (dB) & $\gamma_{\mathrm{opt},n}$ 
& $n$ & $\mathcal{N}(|n\rangle)$ & $\Ccal^2(|n\rangle)$ & $\lambda_n$ & $\Ccal_{\mathrm{opt},\lambda_n}^2$ & $\delta_{\mathrm{opt},n}$ (dB) & $\gamma_{\mathrm{opt},n}$ \\
\hline
0  & 1.00  & 1.00  & 0.00  & 1.00  & 0.00  & 0.000  & 7  & {2.73}  & 15.0  & 10.05 & {14.25} & {11.56} & {0.088}   \\
1  & 1.43  & 3.00  & 1.80  & {2.73}  & {4.38}  & {0.188}  & 8  & 2.87  & 17.0  & 11.30 & {16.01}  & {12.06}  & {0.083}  \\
2  & 1.73  & 5.00  & 3.17  & {4.59}  & {6.63}  & {0.152}  & 9  & 3.02  & 19.0  & 12.72 & {18.02}   & {12.58}  & {0.078}  \\
3  & 1.98  & 7.00  & 4.55  & {6.51}  & {8.15}  & {0.129}  & 10 & 3.15  & 21.0  & 14.01 & {19.84}   & {13.00}  & {0.075}  \\
4  & 2.19  & 9.00  & 5.89  & {8.39}  & {9.25}  & {0.114}  & 11 & 3.28  & 23.0  & 15.37 & {21.76}   & {13.40}  & {0.071}  \\
5  & 2.38  & 11.0  & 7.23  & {10.27} & {10.13} & {0.103}  & 12 & 3.41  & 25.0  & 16.79 & {23.76}   & {13.78} & {0.068} \\ 
6  & 2.56  & 13.0  & 8.62  & {12.23} & {10.89} & {0.095}  & 13 & 3.53  & 27.0  & 18.15 & {25.69}   & {14.12}  & {0.066}  \\
\hline
\end{tabular}
\caption{
\label{table:lambdan} Comparison of the QCS $\mathcal C^2(|n\rangle\langle n|)=2n+1$ of the Fock states with $0\leq n\leq 13$ and, for each $n$, the QCS $\mathcal C^2_{\mathrm{opt},\lambda_n}$ (from Eq.~\eqref{eq:QCSmin}-\eqref{eq:QCSmin2}) of the optimal cubic Gaussian state $\rhoh_{\textrm{opt},\lambda_n}$ with the same Wigner negativity. The values of $\lambda_n$ are defined in Eq.~\eqref{eq:lambdan} and determined numerically; $\delta_{\mathrm{opt}, n}$ and $\gamma_{\mathrm{opt}, n}$ are the corresponding squeezing parameter (dB) and cubicity.  
}
\end{table}
In Table~\ref{table:lambdan}, for $0\leq n\leq 13$, one finds the numerically computed values of the total negativity of $|n\rangle$ and of $\lambda_n$, as well as the values of the QCS $\Ccal^2_{\mathrm{opt}, \lambda_n}$ of the corresponding optimal cubic Gaussian states $\rhoh_{\mathrm{opt}, \lambda_n}$, obtained from Eq.~\eqref{eq:QCSmin2}.  Remark that, in the range of $n$ considered,  the values of $\Ccal^2_{\mathrm{opt}, \lambda_n}$ are close to $\Ccal^2(|n\rangle)=2n+1$, so that the decoherence time of the $n$th Fock state and of the corresponding minimal cubic Gaussian state $\rhoh_{\mathrm{opt}, \lambda_n}$ are comparable.  In addition, in view of Eq.~\eqref{eq:pureQCS}, the mean photon number of $\rhoh_{\mathrm{opt}, \lambda_n}$ is close to $n$. The cubicity $\gamma_{\mathrm{opt}, \lambda_n}$ and squeezing $\delta_{\mathrm{opt}, \lambda_n}$ needed to prepare $\rhoh_{\textrm{opt},n}$ can be read off from
\begin{equation}\label{eq:gammaminn}
\gamma_{\textrm{opt},n}=\frac{\sqrt2}{(1+2\lambda_n^2)^{3/4}}\frac{\lambda_n}{3},\quad \delta_{\textrm{opt}, n}=\frac{8.7}{4}\ln(1+2\lambda_n^2).
\end{equation}
One remarks that the minimal cubicity $\gamma_{\textrm{opt},n}$ needed  decreases with $n$, whereas the squeezing increases; for values of $n$ up to $n\simeq 10$, the  squeezing  remains within current experimental capabilities. Indeed, recent experiments have demonstrated significant progress in generating squeezed states with quantum noise reduction in the range of 5 to 15 dB~\cite{Vahlbruch2008, Mehmet2009, Vahlbruch2016, Shajilal2022, Shen2025}. 
On the other hand, in the optical regime, pure Fock states have so far been generated only up to $n\sim3-5$~\cite{Cooper2013,Afek2010}, with $n\sim10$ reported very recently but at low fidelity~\cite{Bao2025}. Larger pure Fock states ($n\sim100$ or more) have only been achieved in non-optical bosonic platforms, such as superconducting microwave circuits~\cite{Hofheinz2008,Deng2024}.

Whereas the value of $\delta_{\opt,n}$ becomes experimentally inaccessible for large $n$, it is of theoretical interest to understand the asymptotic behaviour of $ \mathcal C^2_{\opt,\lambda_n}$ as a function of $n$.
Recall from Eq.~\eqref{eq:Nlambdaasympt} that for large $\lambda$ one has
\begin{equation}
    \mathcal N_\lambda \simeq 1+c\sqrt\lambda, \quad \textrm{with}\ c\approx 0.745.
\end{equation}
On the other hand, a numerical computation shows that for large $n$ (up to $n=200$)
one has approximately that
\begin{equation}
    \mathcal N(|n\rangle\langle n|)\simeq 1+ c' \sqrt n,\quad \textrm{with}\ c'\approx 0.81.
\end{equation}
This was pointed out already in~\cite{Kenfack_2004}.
Hence
\begin{equation}
    \lambda_n\simeq \frac{\mathcal N_{\lambda_n}^2}{c^2}=\frac{\mathcal N(|n\rangle\langle n|)^2}{c^2}\simeq \left(\frac{c'}{c}\right)^2 n,
\end{equation}
and,
\begin{equation}
    \mathcal C^2_{\opt,\lambda_n}\simeq \sqrt2\lambda_n\simeq \frac1{\sqrt2} \left(\frac{c'}{c}\right)^2\mathcal C^2{(|n\rangle\langle n|)}= 0.836\ C^2{(|n\rangle\langle n|)}.
\end{equation}
This indicates that, for large $n$, the QCS of the optimal cubic Gaussian state with the same Wigner negativity as the $n$th Fock state could be about $80\%$ 
%
of the QCS of {Fock state} $|n\rangle$: its decoherence time would therefore be about $20\%$ larger.
Also, the mean photon number $\langle a^\dagger a\rangle_{\opt,\lambda_n}$ of the optimal cubic Gaussian state with the same Wigner negativity as the $n$th Fock state (computed below, see Eq.~\eqref{eq:momentsoptimalcubicsqueezedvacuum}) satisfies
\begin{equation}
    \langle a^\dagger a\rangle_{\opt,\lambda_n}\simeq 1.04 n,
\end{equation}
asymptotically in $n$. In this sense, optimal cubic Gaussian states provide the same Wigner negativity per photon as Fock states, but they achieve this  with a somewhat longer decoherence time.


\begin{figure}[t!]
\begin{center}
    \includegraphics[trim=0 0 0 0, clip,width=0.8\textwidth]{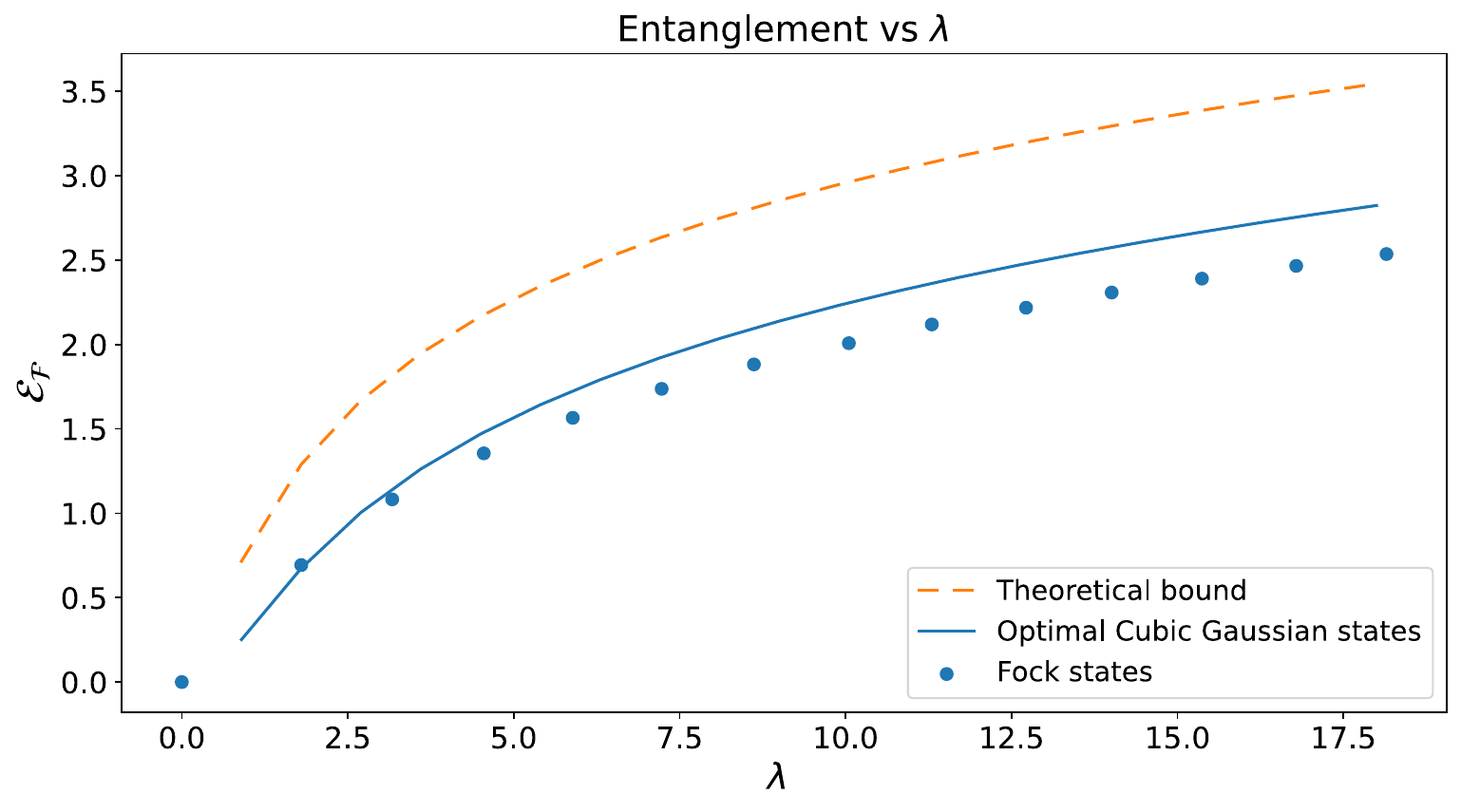}
    \caption{Entanglement generated by optimal cubic Gaussian states and Fock states via 50:50 beam splitter interaction. The dashed orange line refers to the bound introduced in Eq.~\eqref{eq:vN_shannon_upper_bound}.  \label{fig:entanglement_comparison}}
\end{center}
\end{figure}

As a further element of comparison between Fock states and optimal cubic Gaussian states, we consider their capacity for entanglement generation. Entanglement is, like Wigner negativity, a crucial resource in quantum information theory, necessary for claiming a quantum advantage~\cite{RevModPhys.74.197,kenigsberg2006quantum,PhysRevA.101.012349,zhao2025entanglement,PhysRevLett.134.070201}.  In photonic implementations, beam splitter interactions can be used to generate entanglement between input states;  the level of entanglement generated depends on the specific properties of these input states.  As such, states capable of producing high levels of entanglement are therefore more highly sought after~\cite{PhysRevA.75.052106,PhysRevA.89.052302, PhysRevA.54.3824,c9df8bd9-5e34-305f-85e4-f0f511c29050,bruss2002characterizing,Eltschka_2014,makarov2022theory,lhxk-v564}.  Here, the von Neumann-Shannon entropy $\mathcal{E_F} = -\text{Tr}(\rhoh_A \ln \rhoh_A)$, where $\rhoh_A$ is the reduced density matrix of the two-mode pure state $\rhoh$ after the beam splitter, is employed to quantify the entanglement between the two modes of the beam splitter. This analysis is performed via numerical simulation, using the Fock representation of the states with a photon cutoff of 175. 
At different values of $\lambda$, two copies of optimal cubic Gaussian states $\rhoh_{\opt, \lambda}$, rotated by 90 degrees with respect to each other in phase space, are input into the beam splitter. The resulting entanglement between the two output modes is then evaluated. 
As illustrated in Fig.~\ref{fig:entanglement_comparison}, the entanglement generated by the cubic Gaussian states at different values of $\lambda$ is compared with that achieved by two Fock states with the same negativity and interferometer configuration. For larger values of $\lambda$, cubic Gaussian states produce greater entanglement than Fock states, establishing them as a preferable choice when entanglement is the primary criterion for quantum states. We note that the beam splitter alters neither the negativity, nor the QCS of the state.

\begin{figure}[t!]
\begin{center}
    \includegraphics[trim=0 0 0 0, clip,width=0.8\textwidth]{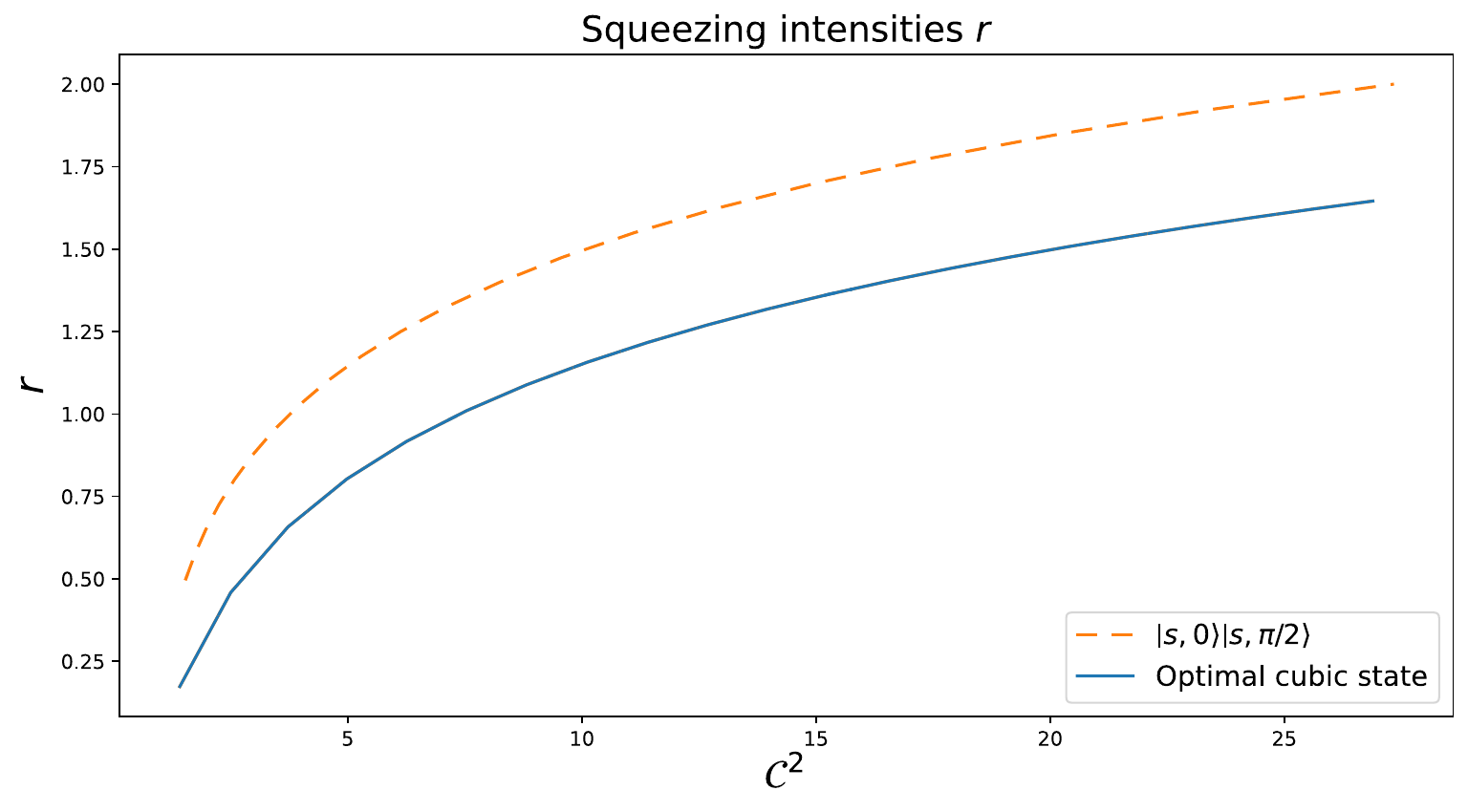}
    \caption{Squeezing intensity $r$ as a function of QCS $\mathcal{C}^2$ for squeezed vacuum states (dashed orange line) and optimal cubic Gaussian states (full blue line).\label{fig:squeezing_comparison}}
\end{center}
\end{figure}

It follows from the bounds obtained in~\cite{hertz_relating_2020} that, when two optimal cubic Gaussian states are presented at the input ports of the beam splitter, as above, then the resulting entropy of formation satisfies:
\begin{equation}\label{eq:vN_shannon_upper_bound}
\mathcal E_{\mathcal F}(\lambda)\leq g(\frac12(\mathcal C^2_{\opt, \lambda}-1)),
\end{equation}
where
\begin{equation}
g(c)=(c+1)\ln(c+1)-c\ln(c).
\end{equation}
The optimal value in the right hand side of Eq. \eqref{eq:vN_shannon_upper_bound} is reached when two squeezed vacua are presented, one rotated with respect to the other by 90 degrees, and with their squeezing adapted so that their QCS equals $C^2_{\opt, \lambda}$~\cite{hertz_relating_2020}. Consequently, for a fixed QCS, squeezed vacuum states would guarantee a higher level of entanglement than the optimal cubic Gaussian states; however, this would require stronger squeezing, as shown in Fig.~\ref{fig:squeezing_comparison}. And, of course, the squeezed vacua don't have any Wigner negativity.  

From the above, we may conclude that the optimal cubic Gaussian states have similar characteristics to the Fock states, from the point of view of their Wigner negativity, of their QCS, of their mean photon number, and of their entanglement generating potential through a beam splitter. 

Of course, one expects the optimal cubic Gaussian states to have a very  different photon number distribution from Fock states, and, actually, to be somewhat similar to squeezed states in that respect.  In the next section, we examine this idea through a comparison of the photon statistics of the optimal cubic Gaussian states with that of their underlying squeezed states. 
In the process, we evaluate several nonclassicality witnesses 
 expressed in terms of moments of the annihilation/creation operators, for those states. 


\section{\texorpdfstring{Moments of $\hat{a}$ and $\hat{a}^\dagger$ in cubic squeezed vacua}{Moments of a and a dagger in cubic squeezed vacua}}
\label{sec:nonclass_photondistr}
To get a better understanding of the cubic Gaussian states, we now   take a closer look at their photon number distribution. We note, however, that we do not have closed-form analytical expressions for
$$
p_n=\langle n|\hat\rho_{\mathrm{G},\gamma}|n\rangle.
$$
This is true even for cubic squeezed vacua: indeed, we do not know of explicit expressions for
\begin{equation}
    c_n(\gamma, r)=\langle n|\Gamma_\gamma S_{r,\pi}|0\rangle,
\end{equation}
because of the presence of the cubic factor. 
Nevertheless, it is straightforward  -- be it a bit tedious -- to obtain  such an expression for the first few moments  of $a$ and $a^\dagger$ in cubic squeezed vacua.  We will see that those allow us to establish the super-Poissonian nature of all cubic Gaussian states through the computation of their Mandel parameter and to study the general shape of their photon number distribution in comparison to that of their underlying squeezed states. We will also establish a sufficient criterion for their nonclassicality using an appropriately chosen moment.  The detailed computations of the moments needed are provided in Appendix~\ref{sec:mandel}.

In quantum optics, states are said to be classical when they are mixtures of coherent states, or, equivalently, when they have a positive Glauber-Sudarshan $P$-function~\cite{titulaer_correlation_1965}. Cubic gaussian states are nonclassical in this sense, since they have a negative Wigner function, which implies that they do not have a positive $P$-function. Similarly, the fact that their  QCS is larger than $1$ provides an independent proof of their nonclassicality~\cite{de_bievre_measuring_2019}, as noted above. 
More generally, the  nonclassicality  of a state can be assessed by evaluating analytically, numerically or experimentally appropriate moments of the creation and annihilation operators in those states. 
Such moments are known to  provide (non-faithful) witnesses of nonclassicality \cite{titulaer_correlation_1965,mandel_sub-poissonian_1979,agarwal_nonclassical_1992,shchukin_nonclassicality_2005,shchukin_nonclassical_2005,arnhem_multicopy_nodate}. Furthermore,   sub-Poissonian photon statistics, characterized by 
\begin{equation}
    (\Delta \hat n)^2 < \langle \hat n\rangle,
\end{equation}
is a feature known to imply nonclassicality. Classical light, on the other hand, is known to be super-Poissonian, meaning that 
$
    (\Delta \hat n)^2 \geq \langle \hat n\rangle.
$
But nonclassical light can also be super-Poissonian, with squeezed vacua, known to be nonclassical, being an example. The non-Poissonian nature of quantum light is typically assessed using the Mandel Q parameter:
\begin{equation}
    Q = \frac{\mean{\hatdn{a}{2} \hatn{a}{2}} - \mean{\hatd{a} \hat{a}}^2}{\mean{\hatd{a} \hat{a}}}.
\end{equation}
One has 
\begin{equation}
    Q\geq 0 \Leftrightarrow (\Delta \hat n)^2 \geq\langle \hat n\rangle, 
\end{equation}
so that positivity of the Mandel parameter is equivalent to super-Poissonian statistics, referred to also as ``bunching''. Indeed, the Mandel parameter is related to the intensity autocorrelation at zero delay $g^{(2)}(0)$: 
\begin{equation}
    g^{(2)}(0)=1+\frac{Q}{\langle \hat n\rangle}=\frac{\langle {a^\dagger}^2a^2\rangle}{\langle a^\dagger a\rangle^2},
\end{equation}
and $g^{(2)}(0)>1$ is known as ``bunching'', $g^{(2)}(0)<1$ as ``anti-bunching''. 
Note that $g_2(0)>1$ or $Q\geq 0$ is equivalent to 
\begin{equation}
    \frac{\Delta\hat n}{\langle\hat n\rangle}\geq \langle \hat n\rangle^{-1/2}.
\end{equation}
The quantity  $\frac{\Delta\hat n}{\langle\hat n\rangle}$ has a direct geometric meaning for the graph of the photon number distribution: it measures its relative spread around its mean. This means in particular that, if this relative spread is large, then the distribution shows bunching. Note that, if $\frac{\Delta\hat n}{\langle\hat n\rangle}\geq 1$, then one expects a large contribution of low lying photon numbers, and of the vacuum, to the photon number distribution. 

 To capture the possible nonclassical nature of super-Poissonian states, one can use other moments of $a$ and $a^\dagger$. In fact, as seen in Eq.~\eqref{eq:pureQCS} the QCS 
is such a moment expression for pure states, and it does furnish a faithful nonclassicality witness for pure states. As a further example, in~\cite{arnhem_multicopy_nodate}, it is shown that, if 
\begin{equation}
    d_{23}=\langle \hatd{a} \hat{a} \rangle^2-\langle \hatn{a}{2}\rangle\langle \hatdn{a}{2}\rangle<0,
\end{equation}
then the state is nonclassical. This criterium provides a (non faithful) nonclassicality witness that detects the nonclassicality of squeezed states, for example. $d_{23}$ is  interesting since, like the QCS, it can be directly measured using two copies of the state, a beam splitter and photon counting, see~\cite{arnhem_multicopy_nodate, griffet_interferometric_2023}.   

We first consider cubic vacuum states, for which one finds
\begin{equation}\label{eq:mandelcubicvacuum}
   \langle a^\dagger a\rangle_\gamma=\frac{27}{8}\gamma^2, \quad \Delta^2_\gamma(a^\dagger a)=\frac{1}{8}\left(63\gamma^2+ 972\gamma^4\right),\quad  Q_\gamma
    = \frac{4}{3} + 36 \gamma^2, \quad d_{23}=0.
\end{equation}
 Since the $Q$-parameter is positive, one concludes that cubic vacua are necessarily super-Poissonian. Since $d_{23}=0$,  $d_{23}$ does not detect the nonclassicality of cubic vacua, either, contrary  to the QCS. In addition, for large cubicity $\gamma$ one observes that both the mean and the standard deviation of their photon number distribution grow quadratically in $\gamma$, so that
\begin{equation}\label{eq:spread_cubicgaussian}
    \Delta_\gamma(a^\dagger a)\simeq 
    4 \sqrt{\frac{2}{3}} \langle a^\dagger a\rangle_\gamma \simeq 3.27 \langle a^\dagger a\rangle_\gamma .
\end{equation}
As a result, one expects the support of these photon number distributions to be large and  stretch from $n=0$ to $n\sim \langle a^\dagger a\rangle_\gamma+ 3\Delta_\gamma(a^\dagger a)\sim  10.80 \langle a^\dagger a\rangle_\gamma$. By comparison, for strongly squeezed states, one has
\begin{equation}\label{eq:spread_squ}
   \Delta_r(a^\dagger a) \simeq \sqrt2\langle a^\dagger a\rangle_r.
\end{equation}
In other words, the relative spread of the photon distribution of a cubic vacuum state (with large $\gamma$) around its mean is approximately twice that of a squeezed state (with large $r$). 

For general  cubic squeezed Gaussian states, we find:
\begin{equation}\label{eq:photonmeannumbercubicsqueezed}
    \mean{\hat{a}^\dagger \hat{a}}_{\gamma, r, \pi} = \frac{D(r,\gamma)}{8},
   \quad \Delta^2_{\gamma, r,\pi}(\hat a^\dagger \hat a)= \frac{A(r)\, \gamma^{4}
+ B(r)\, \gamma^{2} + C(r) + D(r, \gamma)}{8}, 
\end{equation}
with
\begin{eqnarray}
    A(r) &=& 972 e^{8r} , \quad\quad\quad\quad\quad\quad\quad\quad\quad\quad\quad\quad
    B(r) = 18 \, e^{4r} \left(2 + 7 \sinh^2(r) + 5\cosh(r)\sinh(r)\right)  , \\
    C(r) &=&  8 \sinh^2(r)\left(\cosh^{2}(r)+ \sinh^2(r)\right) , \quad
    D(r, \gamma) = 27 \gamma^2 e^{4 r}  + 8 \sinh^2{(r)}.
\end{eqnarray}
This yields, for the Mandel parameter
\begin{equation}\label{eq:Q_cubicsqueezed}
    Q_{\gamma, r, \pi}=\frac{A(r)\, \gamma^{4}
+ B(r)\, \gamma^{2} + C(r)}{ D(r, \gamma)},
\end{equation}
which is seen to be positive, so that the cubic Gaussian states are all super-Poissonian. Their nonclassical nature is not detected by the Mandel parameter: this is not too surprising since it does not detect the nonclassicality of cubic vacua, nor of squeezed states. 

 The expressions obtained in Eq.~\eqref{eq:photonmeannumbercubicsqueezed} and in Eq.~\eqref{eq:Q_cubicsqueezed} are explicit but unwieldy. It is instructive, on the other hand, to analyze them for the case of the optimal cubic Gaussian state, in function of $\lambda$. We specifically study the relation between the mean photon number and its standard deviation for optimal cubic Gaussian states, when  $\lambda$ is large. This will provide us with qualitative information about the general shape of the photon number distribution of those states.  One finds readily that 
\begin{eqnarray}
    A(r_\textrm{opt}) \gamma_\opt^4 &\simeq& 24 \lambda^2 , \quad
    B(r_\textrm{opt}) \gamma_\opt^2 \simeq 2 \lambda (6 \lambda + 2 \, \sqrt{2})  , \\
    C(r_\textrm{opt}) &\simeq&  2 \lambda^2 , \quad
    D(r_\textrm{opt}, \gamma_\opt) \simeq 5 \, \sqrt{2} \lambda.
\end{eqnarray}
Hence 
\begin{equation}\label{eq:momentsoptimalcubicsqueezedvacuum}
    \langle a^\dagger a\rangle_{\opt}\simeq \frac5{8}\sqrt2\lambda, \quad \Delta^2_{\textrm{opt}}(a^\dagger a)\simeq \frac{38 \lambda^2 + 9 \sqrt{2} \lambda}{8}, \quad Q_{\textrm{opt}}\simeq \frac{38}{5 \sqrt{2}} \lambda + \frac{4}{5}.
\end{equation}
Consequently, for large $\lambda$, one finds
\begin{equation}\label{eq:spread_min}
    \Delta_{\opt}(a^\dagger a)\simeq 2.47 \langle a^\dagger a\rangle_{\opt}.
\end{equation}
This is to be compared to Eq.~\eqref{eq:spread_cubicgaussian}
 and Eq.~\eqref{eq:spread_squ}: the relative spread of the photon number distribution of the optimal cubic Gaussian states around their mean is between that of a squeezed state and that of a cubic vacuum state.
 \begin{figure}
    \centering
     \subfloat[\label{fig:photon_distribution_relative}]
    {\includegraphics[width=0.45\linewidth]{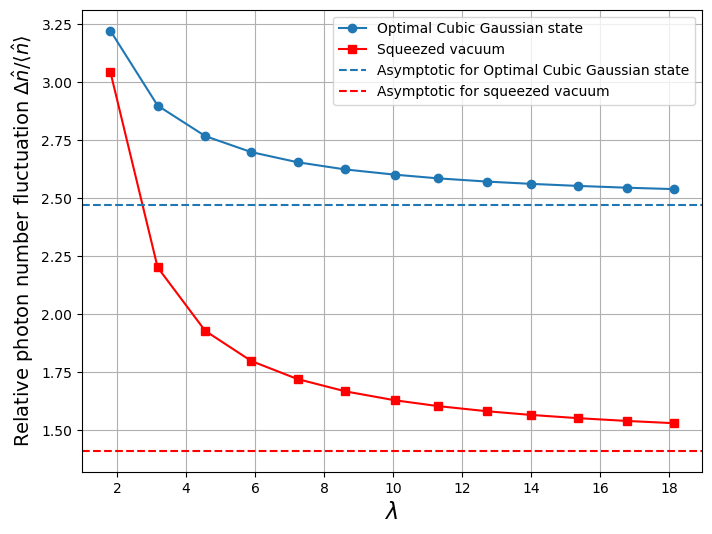}}
    \hspace{0.1cm}
     \subfloat[\label{fig:photon_distribution_mean}] 
    {\includegraphics[width=0.45\linewidth]{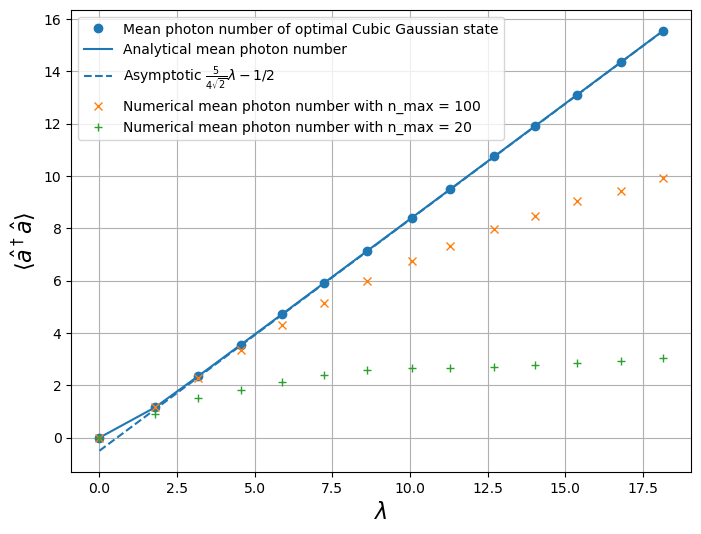}}
    \caption{(a) Plot of the photon number standard deviation $\Delta_{\opt}\hat n$ relative to the mean photon number $\langle \hat n\rangle_{\opt}$ of the optimal cubic Gaussian states, as a function of $\lambda$ (blue dots), obtained from the analytic formulas in the text. Same, for the squeezed vacuum with $r=r_{\opt}$ (red squares).  The dashed lines indicate the asymptotic values obtained in~\eqref{eq:spread_min} and~\eqref{eq:spread_squ}. (b) The mean photon number of optimal cubic Gaussian states as a function of $\lambda$, obtained from the analytical formulas in the text (blue line) and computed numerically (blue dots). The (green) dashed line  plots the asymptotic formula in Eq.~\eqref{eq:momentsoptimalcubicsqueezedvacuum}.  Crosses (green and orange) indicate numerically computed values of the mean photon number of the same states, but with the sum truncated a $n_{\max}$ as indicated. See the main text for further comment.  }
    \label{fig:photon_distribution_spread}
\end{figure}
In Fig.~\ref{fig:photon_distribution_spread} we plot $\Delta_{\opt}(a^\dagger a)/\langle a^\dagger a\rangle_{\opt}$ for the optimal cubic Gaussian states, as well as $\Delta_{\mathrm r}(a^\dagger a)/\langle a^\dagger a\rangle_{\mathrm r}$, with $r=r_{\opt}$, as a function of $\lambda$. One notices that the cubic gate with cubicity $\gamma_{\opt}$ considerably enhances the relative spread of the photon distribution of the squeezed vacuum with squeezing parameter $r_{\mathrm{opt}}$ around its mean, even at very large $\lambda$, where the cubicity $\gamma_{\mathrm{opt}}$ is nevertheless small, as we saw. In fact, comparing Eq.~\eqref{eq:spread_min} to Eq.~\eqref{eq:spread_squ}, one notices that the relative   standard deviation is increased by $75\%$, asymptotically. Note that the small cubicity applied at large $\lambda$ also considerably enhances the mean photon number: one easily checks that 
\begin{equation}
\langle a^\dagger a\rangle_{\mathrm{opt}}\simeq 2.5 \langle a^\dagger a\rangle_{r_{\mathrm{opt}}},
\end{equation}
which constitutes a 150\% increase in the mean photon number.

\begin{figure}[t!]
  {\includegraphics[width=17 cm]{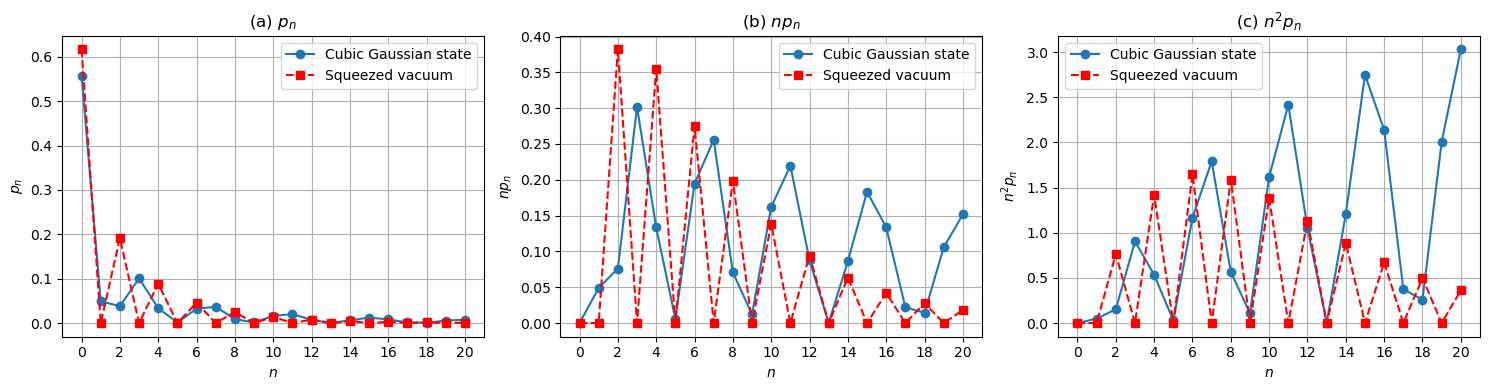}}
  \hspace{0.1cm}
   \caption{Left panel. Numerically computed photon number distribution $p_n$ up to $n = 20$ of the optimal cubic squeezed state (blue dots) and of the underlying squeezed vacuum state (red squares) for $\lambda = \lambda_4 = 5.89$. The corresponding value of  squeezing is $r_{\mathrm{opt}} = 1.06$ and of cubicity is $\gamma_{\mathrm{opt}} = 0.11$.
   The middle panel displays $np_n$ statistics and the right panel displays $n^2 p_n$ statistics.
}\label{fig:Pn_cubic}
\end{figure}

The asymptotic behaviour in $\lambda$ of the various moments that we derive here tends to be very accurate, even at rather low values of $\lambda$. To illustrate this observation, we show in the second panel of Fig.~\ref{fig:photon_distribution_spread}  the asymptotic expression of the mean photon number $\langle a^\dagger a\rangle_{\mathrm{opt}}$ as a function of $\lambda$ (blue dashed line) alongside its analytical expression (blue line): the agreement is seen to be excellent down to $\lambda\simeq 2$. 

We further point out that, to obtain the same accuracy from a numerical computation of the $p_n$, one has to compute the $p_n$ for very large values of $n$. To illustrate this, we plotted $\sum_{n=1}^{n_{\max}} np_n$ in Fig. \ref{fig:photon_distribution_mean} for $n_{\max}=20$ (green cross markers) and $n_{\max}=100$ (orange diagonal cross markers). One notices the poor approximation of the analytical expression by these quantities. The phenomenon is enhanced when considering the higher moments needed to determine $Q$ (not shown). The accurate numerics shown in the figure is obtained by adapting $n_{\max}$ to $\lambda$, as $\lambda$ grows: for example, we need to take $n_{\max}=5000$ for $\lambda=20$. The origin of this discrepancy can be read off from Fig.~\ref{fig:Pn_cubic}. Panel (a) shows a global shape of the photon distribution of the optimal cubic squeezed state (blue line) that looks superficially similar to the one of the underlying squeezed state (red line), even though, as expected, the  $p_{2n+1}$  do no longer vanish.   Nevertheless, it can indeed be observed in panels (b) and (c) of Fig.~\ref{fig:Pn_cubic} that  $np_n$ and $n^2p_n$ remain non-negligible for much larger values of $n$ for the optimal cubic squeezed state than for the underlying squeezed state. This is a consequence of the larger spread of those photon number distributions as witnessed by Eq.~\eqref{eq:spread_squ}, Eq.~\eqref{eq:momentsoptimalcubicsqueezedvacuum} and Eq.~\eqref{eq:spread_min} and that is not apparent to the naked eye in panel (a). 

We conclude from this analysis that it is not feasible to determine the mean photon number or its standard deviation for cubic squeezed states from measuring their photon number statistics and from a post-treatment of the data. Indeed, this would mean measuring photon number statistics accurately to high values of $n$. Note however that the mean photon number and its variance can be measured directly from balanced homodyne detection, via moment tomography of the quadratures, without any need to reconstruct the full photon-number distribution $p_n$. This means that the formulas above provide an immediate and practical way to assess the nonclassical and nongaussian character of the cubic squeezed vacua.

We finally consider the $d_{23}$ non-classicality criteria for cubic Gaussian states. From Eq.~\eqref{eq:photonmeannumbercubicsqueezed}, as well as Eq.~\eqref{eq:a2_cubic}, one readily computes 
\begin{equation}
    d_{23}=-\sinh{r}(\sinh{r}-6\tau \exp(r)), \quad \tau=\frac{9}{8}\gamma^2\exp(4r).
\end{equation}
Therefore, 
\begin{equation}
    d_{23}<0 \Leftrightarrow \gamma^2< \frac2{27}\exp(-4r)(1-\exp(-2r)). 
\end{equation}
Hence $d_{23}$ detects the nonclassicality of a cubic squeezed state provided the cubicity $\gamma$ is not too large compared to the squeezing. In particular, it does not detect the nonclassicality of a cubic vacuum ($r=0$), and does detect the nonclassicality of a squeezed state ($\gamma=0$), as we already noticed.  One remarks that this expression is considerably simpler than the one for the Mandel parameter or even for the QCS, for the same states. 
Using Eq.~\eqref{eq:squeezingmin} and Eq.~\eqref{eq:gammamin}, one however readily checks that this condition is not satisfied for optimal cubic Gaussian states so that $d_{23}$ does not detect their nonclassicality. This is a consequence of the fact that, although, for large $\lambda$, the cubicity $\gamma_{\opt}$ becomes small, it is not small enough compared to the growing squeezing $r_{\opt}$ for that to be the case.

Higher moments of $a$ and $a^\dagger$ can in principle also be computed for the cubic squeezed vacua, and in this manner all moment-based nonclassicality criteria can be  evaluated. However, the computations become increasingly cumbersome and the results are not enlightening. 


\section{Thermal noise induced negativity loss}\label{sec:thermal}

\begin{figure}[t!]
  \subfloat[\label{fig:Neg_vs_q}]{\includegraphics[width=5.5cm]{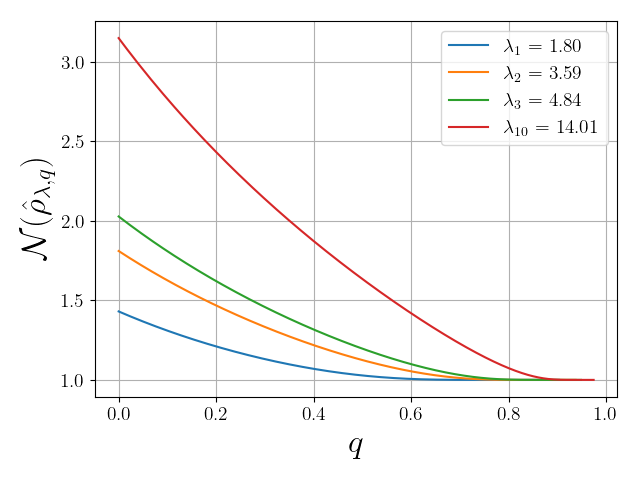}}
  \hspace{0.1cm}
  \subfloat[\label{fig:C2_vs_q}]{\includegraphics[width=5.5
  cm]{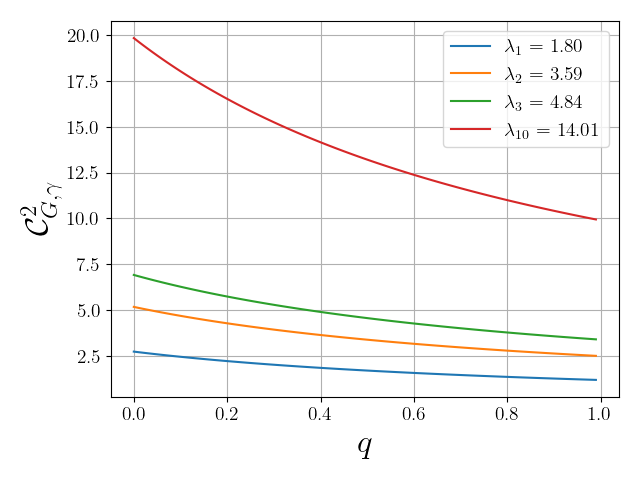}}
  \hspace{0.1cm}
  \subfloat[\label{fig:q_star}]{\includegraphics[width=5.5cm]{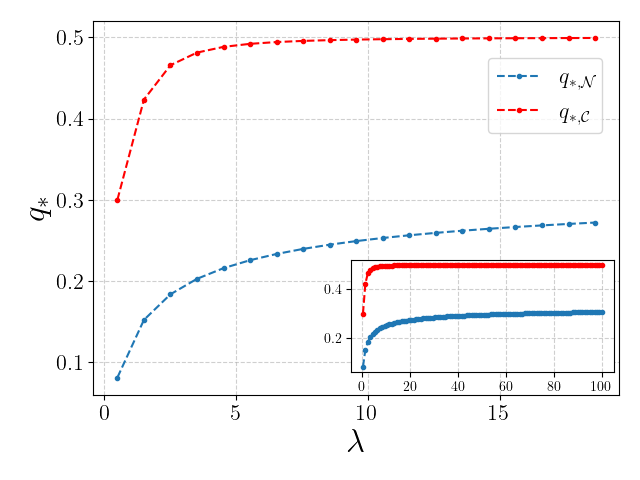}}
   \caption{(a) Wigner negativity of the noisy optimal cubic Gaussian states $\hat\rho_{\lambda,q}$ as a function of the noise parameter $q$, for values of $\lambda$ as indicated; (b) QCS of the same states versus noise; (c)  The red curve for $q_{*,\mathcal{C}}$ is the analytic plot from Eq.~\eqref{eq:qcsthermalhalflife} evaluated for each $\lambda_n$ while the blue curve of $q_{*,\mathcal{N}}$ is computed numerically for each $\lambda_n$;   $q_{*,\mathcal{N}}$ is defined in \eqref{eq:qstarN}. \label{fig:noise} 
   }
\end{figure}

{In this section, we briefly analyse how replacing the vacuum with a thermal input state affects the optimal cubic Gaussian states introduced in Section ~\ref{s:optimalCGstates}. This thermal input substitution allows us to test the robustness of the optimized squeezing and cubic phase gate operations against realistic noise.}
For that purpose, we consider, for $\lambda\geq 1/\sqrt2$ and $0\leq q<1$, 
\begin{equation}
    \hat\rho_{\lambda, q}=\Gamma_{\gamma_{\opt}}S_{\zeta_{\opt}} \hat\rho_{\textrm{Th}}S_{\zeta_{\opt}}^\dagger \Gamma_{\gamma_{\opt}}^\dagger.
\end{equation}
Here $\hat\rho_{\textrm{Th}}$ is the thermal state defined in Eq.~\eqref{eq:SqTh} and $\gamma_{\opt}$ and $\zeta_{\opt}$ depend on $\lambda$ as in Eq.(\ref{eq:gammamin}) and Eq.(\ref{eq:squeezingmin}), so that $\hat\rho_{\lambda, 0}=\hat\rho_{\opt, \lambda}$. One then has 
\begin{equation}
\mathcal N(\hat\rho_{\lambda, q})=\mathcal N_{\Lambda(\lambda, q)},
\end{equation}
where
\begin{equation}
\begin{split}
\Lambda(\lambda, q)&= {\frac{3}{\sqrt{2}}\gamma_{\opt}(\lambda)\left(\frac{1-q}{1+q}\right)^{3/2}\exp(3r_{\opt}(\lambda))=\left(\frac{1-q}{1+q}\right)^{3/2}\lambda.}
\end{split}
\end{equation}
Panel~(a) of Fig.~\ref{fig:noise} shows the effect of the noise parameter $q$ on the Wigner negativity $\mathcal N(\hat\rho_{\lambda,q})$, for various values of $\lambda$. To obtain a quantitative measure of the negativity loss, we proceed as follows. We define the relative negativity loss $\eta(q)$ as follows
\begin{equation}\label{eq:negloss}
\eta(q)=\frac{\mid \mathcal N(\hat\rho_{\lambda,q})-\mathcal N(\hat\rho_{\lambda,0})\mid}{\mathcal N(\hat \rho_{\lambda,0})-1}\simeq\frac{\mid \frac{\rd \mathcal N_{\Lambda(\lambda,0)}}{\rd q}\mid}{\mathcal N(\hat \rho_{\lambda,0})-1}q.
\end{equation}

Hence {$q_{*,\mathcal N}$,} defined via 
\begin{equation}\label{eq:qstarN}
   \eta(q_{*,\mathcal N})=\frac12, 
\end{equation}
is the noise-level at which half the initial Wigner negativity is lost.  We will use $q_{*,\mathcal N}$ as a measure of the noise-sensitivity of the Wigner negativity. Panel (c) of Fig.~\ref{fig:noise}  shows the dependence of $q_{*,\mathcal N}$ as a function of $\lambda$. One notices that it increases with $\lambda$, showing a diminishing of the noise sensitivity with increasing Wigner negativity. For large values of negativity (and hence large values of $\lambda$), $q_{*,\mathcal N}$ can be estimated as follows. One has that 
$$
\frac{\rd\mathcal N(\hat\rho_{\lambda, q})}{\rd q}=\frac{\rd\ }{\rd q}\mathcal N_{\Lambda(\lambda, q)}\simeq-\frac{3c}{2}\sqrt\lambda,
$$
so that, for large $\lambda$, $q_{*,\mathcal N}$ saturates:
{\begin{equation}\label{eq:qstarNasymp}
q_{*,\mathcal N}\simeq \frac13.
\end{equation}}
This corresponds to a thermal input state with a mean photon number equal to $0.5$.
One concludes that the Wigner negativity of the optimal cubic Gaussian state is strongly noise dependent, even at large values of Wigner negativity, where is attains its minimal value.

Similarly,  from Eq.(\ref{eq:QCSGaussianGamma}) we can have a quantitative measure of the QCS loss 
\begin{equation}
\mathcal{C}^2_{\mathrm{opt},\lambda,q}=
A\left(\frac{1-q}{1+q}\right)+ B \lambda^2,
\end{equation}
with $A=\cosh(2 r_{\mathrm{opt}})$, $B=\left(\mathrm{e}^{-2r_{\mathrm{opt}}}\cos^2(\phi/2)+\mathrm{e}^{2r_{\mathrm{opt}}}\sin^2(\phi/2)\right)^{-1}$ and $\phi=\pi$.
Therefore,
\begin{equation}
\left.\frac{d \mathcal{C}^2_{\mathrm{opt},\lambda,q}}{dq}\right|_{q=0}=
-2A=-2\cosh\left(\frac{1}{2}\ln\left(1+2\lambda^2\right)\right).
\end{equation}
Then
\begin{equation}
\eta(q;\mathcal{C})
=
\frac{\left|\mathcal{C}^2_{\mathrm{opt},\lambda,q}-\mathcal{C}^2_{\mathrm{opt},\lambda,q=0}\right|}{\mathcal{C}^2_{\mathrm{opt},\lambda,q=0}}
\approx
\frac{\left|\left(\frac{d \mathcal{C}^2_{\mathrm{opt},\lambda,q}}{dq}\right)_{q=0}\right|}{\mathcal{C}^2_{\mathrm{opt},\lambda,q=0}}q
=
\frac{2 A}{A+B\lambda^2}q .
\end{equation}
Let us introduce $q_{*,\mathcal{C}}$, defined via $\eta(q_*;\mathcal{C}) = 1/2$ which will be the noise level that causes one looses half the QCS:
\begin{equation}
q_{*,\mathcal{C}}
=
\frac{A+B\lambda^2}{4A}
=
\frac{\cosh(2 r_{\mathrm{opt}})+\mathrm{e}^{-2 r_{\mathrm{opt}}}\lambda^2}{4\cosh(2 r_{\mathrm{opt}})}.
\end{equation}
We can replace $r_\mathrm{opt}$ with its expression in terms of $\lambda$ according to Eq.(\ref{eq:squeezingmin}), and one finds: 
\begin{equation}\label{eq:qcsthermalhalflife}
   q_{*,\mathcal{C}}
= \frac{1}{4}\frac{(1+2\lambda^2)}{(1+\lambda^2)},
\end{equation}
which saturate to $1/2$ for large $\lambda$ in agreement with the numerical simulation displayed on Fig. \ref{fig:q_star}.

In the next section we  provide a second application of our findings on the relation between the Wigner negativity and the QCS of cubic Gaussian states to the approximation of ideal cubic phase states by (optimal) cubic Gaussian states. 



\section{Approximating the ideal cubic phase states}\label{sec:approx_ideal_cubic}
In \cite{gottesman_encoding_2001}, the so-called ideal cubic phase state $|\gamma, p=0\rangle$ is defined as
\begin{equation}\label{eq:cubicideal1}
    |\gamma, p=0\rangle =\frac1{\sqrt{2\pi}}\int  \exp(-i\gamma x^3)|x\rangle 
 \rd x=\Gamma_\gamma|p=0\rangle,
\end{equation}
where 
$|p=0\rangle$ is the generalized momentum eigenstate with zero momentum. The ideal cubic phase states have been proposed as a tool to implement the cubic phase gate \cite{gottesman_encoding_2001} which is a non-Gaussian unitary that can be used to complete the set of Clifford gates for universal quantum computing. The ideal cubic phase state is the nonnormalizable eigenstate of $\hat p+3\gamma\hat q^2$ with eigenvalue $0$. 
Since, for example, it  has infinite photon number and infinite Wigner negativity, it is  not physical.
From Eq.~\eqref{eq:WignerLambda}, one finds the Wigner function of the ideal cubic phase state 
\begin{equation}\label{eq:cubicideal2}
|\gamma, p=0\rangle\simeq\lim_{r\to+\infty} \Gamma_\gamma S_{-r}|0\rangle,
\end{equation}
in the limit of large squeezing: 
\begin{equation}
    W_{|\gamma,p=0\rangle} (q, p) = 2 \left|\frac{4}{3 \gamma}\right|^{1/3} \textrm{Ai}\left( \left( \frac{4}{3 \gamma}\right)^{1/3} (3 \gamma q^2 + p) \right).
\end{equation}
It is  not normalized and the  formula makes the infinite negativity of the ideal cubic state clear.

To better understand in which sense the cubic squeezed vacua approximate the ideal cubic phase state, we proceed as follows.  For any pure state $|\psi\rangle$, we have
\begin{equation}
\Gamma_\gamma|\psi\rangle=\int \rd x \exp(-i\gamma x^3) \langle x|\psi\rangle |x\rangle.
\end{equation}
Hence
\begin{equation}\label{eq:cubicsqueezed}
\Gamma_\gamma S_{-r}|0\rangle=\int \rd x \exp(-i\gamma x^3) \langle x|S_{-r}|0\rangle |x\rangle,\quad{\text{with}\quad \langle x|S_{-r}|0\rangle = \frac{\e^{-\frac{r}2}}{(\pi)^{1/4}}\exp(-\frac12 \e^{-2r}x^2)).}
\end{equation}
Comparing this last equation to Eq.~\eqref{eq:cubicideal1}, one sees that the Gaussian factor in Eq.~\eqref{eq:cubicsqueezed} regularizes the integral on an effective length scale of size $\e^r$.
This justifies the suggestion that cubic squeezed vacua can be used to approximate the ideal cubic phase state 
\cite{gottesman_encoding_2001,ghose_non-gaussian_2007,arzani_polynomial_2017,takagi_convex_2018,yukawa_emulating_2013,zheng_gaussian_2021, yanagimoto_engineering_2020, eriksson_universal_2023, kala_cubic_2022, kala:25}.
We have, in fact, for any cubic Gaussian state, that 
\begin{equation}\label{eq:meanvariance}
\begin{split}
    \tr (( \hat{p} + 3 \gamma \hat{q}^2) \rhoh_{G,\gamma}) &= \mean{\hat{p}}_{G} = p_0, \\
     \tr (( \hat{p} + 3 \gamma \hat{q}^2)^2 \rhoh_{G,\gamma}) &= \mean{\hat{p}^2}_{G}=\sigma_{pp}+p_0^2,
\end{split}
\end{equation}
so that the variance of the operator $\hat{p} + 3 \gamma \hat{q}^2$ for a cubic Gaussian state is equal to the variance of the $\hat p$-quadrature of the initial Gaussian state $\rhoG$. In that sense, one may consider that, in the limit of infinite squeezing in the $\hat p$-quadrature of the cubic squeezed vacuum state (for which $p_0=0$), at fixed cubicity $\gamma$, one approaches the ideal cubic phase state, which is an eigenstate of $(\hat p+3\gamma \hat p^2)$ and has, in that sense, vanishing variance.
 In the same regime of large $r$ and fixed cubicity $\gamma$, $\lambda$ is large and hence so is the Wigner negativity of these states. In that limit, however, the QCS grows also beyond any bound, and consequently the decoherence time of these states becomes arbitrarily small, at fixed cubicity. This is not desirable.  Another approach could be the following.  One could decide to try and minimize $\sigma_{pp}$ at fixed Wigner negativity, thereby minimizing the variance of $(\hat p+3\gamma \hat p^2)$. As we now show, this is not possible, since $\sigma_{pp}$ can be taken arbitrarily small, while keeping the Wigner negativity fixed, and that will inevitably drive the QCS to very large values and hence again lead to a very short decoherence time. Indeed, for cubic squeezed vacua the quadrature coherent scale (see Eq.~\eqref{eq:QCScubiccovelements}) is 
\begin{eqnarray}
    \mathcal{C}_{r, \gamma}^2 = \frac{1}{\det V} \left( \sigma_{qq} + \sigma_{pp} + 18 \gamma^2 \sigma_{qq}^2 \right),
\end{eqnarray}
where : $\det V = 4 ~\sigma_{qq} \sigma_{pp} = 1$, as the cubic squeezed state is a pure state.
At fixed Wigner negativity, one can substitute $\gamma$ for $\lambda$ according to Eq.~\eqref{eq:lambdadef}, such that the QCS equals
\begin{eqnarray}
    \mathcal{C}_{r, \gamma}^2 = \left( \frac{1}{4 \sigma_{pp}} + (1 + 8 \lambda^2) \sigma_{pp} \right).
\end{eqnarray}
Hence, by taking $\sigma_{pp}$ arbitrarily small, the QCS will become arbitrarily large and hence the decoherence time of the state arbitrarily short. Note that, in this process, $\gamma$ tends to zero as well. 

The optimal cubic Gaussian states we constructed, on the other hand,  minimize the QCS, and therefore maximize the coherence time, at fixed Wigner negativity $\mathcal N_\lambda$. This minimum is reached at a finite value of $\sigma_{pp}$, which becomes smaller as the negativity increases. Precisely, we have 
\begin{equation}
\sigma_{pp,\mathrm{opt}}=\frac12\exp(-2r_{\mathrm{opt}})\simeq \frac1{2\sqrt2}\lambda^{-1}\simeq \frac1{2\sqrt2}c^2\mathcal N_\lambda^{-2},\quad \gamma_{\mathrm{opt}}\simeq \frac1{3\,2^{1/4}}\lambda^{-\frac{1}{2}}\simeq \frac1{3\,2^{1/4}}c\mathcal N_\lambda^{-1}.
\end{equation}
In conclusion, one finds 
\begin{equation}
    \langle (\hat p+3\gamma_{\mathrm{opt}}\hat q^2)^2\rangle_{\mathrm{opt},\lambda}\simeq 
    \frac1{2\sqrt2}c^2\mathcal N_\lambda^{-2}\simeq 0.20 \mathcal N_{\lambda}^{-2}.
\end{equation}

The authors of~\cite{kala_cubic_2022,kala:25} propose to measure the quality of the approximation of the ideal cubic state by a state $\rhoh$ by the so-called nonlinear squeezing, which is defined as the relative variance
\begin{equation}
\xi_3(\gamma)=\frac{\mathrm{Var}_{\rhoh}(\hat p+3\gamma \hat q^2)}{\min_{\mathrm \rhoG} \mathrm{Var}_{\rhoG}(\hat p+3\gamma \hat q^2)}.
\end{equation}
\begin{table}[t!]
\centering
\begin{tabular}{|c|c|c|c|c|c|c|c||c|c|c|c|c|c|}
\hline
Fig. & $\gamma$ & $r$ & $\delta$ (dB) & $\sigma_{pp}$ & $\Ccal^2$ & $\lambda$ & $\mathcal N_\lambda$ & $\gamma_{\opt}$ & $r_{\opt}$ & $\delta_{\opt}$ (dB) & $q_{\mathrm{min}}$ & $\sigma_{pp, \opt}$ & $\Ccal_{\opt}^2(\lambda)$ \\
\hline 
(a) & 0.0551 & 0.576 & 5.00 & 0.158 & 1.88 & 0.658 & 1.11 & 0.195 & 0.173 & 1.51 & 0.018 & 0.353 & 1.41 \\
(b) & 0.11 & 0.610 & 5.30 & 0.148 & 2.47 & 1.45 & 1.34 & 0.198 & 0.412 & 3.59 & 0.000 & 0.219 & 2.28 \\
(c) & 0.1 & 0.690 & 5.99 & 0.126 & 2.82 & 1.68 & 1.41 & 0.191 & 0.473 & 4.12 & 0.000 & 0.194 & 2.58 \\
(d) & 0.3 & 0.500 & 4.34 & 0.184 & 4.54 & 2.85 & 1.67 & 0.159 & 0.712 & 6.20 & 0.000 & 0.120 & 4.16 \\
(e) & 0.233 & 0.693 & 6.02 & 0.125 & 6.03 & 3.95 & 1.88 & 0.138 & 0.868 & 7.55 & 0.000 & 0.088 & 5.68 \\
\hline
\end{tabular}
 \caption{\label{table:ideal}
 {The second and third column show the cubicity and squeezing parameters from \cite{zheng_gaussian_2021}, \cite{eriksson_universal_2023}, \cite{yanagimoto_engineering_2020}, \cite{kudra_robust_2022} and \cite{kala_cubic_2022} labeled respectively by (a), (b), (c), (d) and (e) in the first column. The fourth column is the squeezing expressed in dB. The fifth column show the variance in the $\hat p$-quadrature of the underlying squeezed state. The sixth column show the QCS computed with the parameters $\gamma$ and $\zeta$. In the seventh and eighth column are displayed the values for the parameter $\lambda$ and the negativity of the cubic Gaussian states whose  Wigner functions of (a), (b), (c) and (e) are shown in Fig.~\ref{fig:cubicphasestate_examples}. The remaining columns on the right show the optimal cubic Gaussian state parameters $(\gamma_{\mathrm{min}}, r_{\mathrm{min}}, q_{\mathrm{min}})$, $\hat p$-quadrature variance and QCS with the same Wigner-negativity $ \mathcal N_\lambda$.}}
\end{table}
One says that the state $\rhoh$ is nonlinearly squeezed if $\xi_3(\gamma)<1$, wich means that the variance of $\hat p+3\gamma\hat q^2$ is smaller than the best variance that can be obtained with a (convex mixture of) Gaussian state. Clearly, $\xi_3(\gamma)$ is a witness of nongaussianity, since $\xi_3(\gamma) \geq 1$ for Gaussian states. States for which $\xi_3(\gamma)<1$ are said to be nonlinearly squeezed: the smaller $\xi_3(\gamma)$, the better the approximation.  The denominator of $\xi(\gamma)$ is explicitly determined in~\cite{kala_cubic_2022}:
\begin{equation}
    \min_{\mathrm \rhoG} \mathrm{Var}_{\rhoG}(\hat p+3\gamma \hat q^2)=3\left(\frac12\right)^{5/3}|3\gamma|^{\frac23}.
\end{equation}
 For the optimal cubic Gaussian states, one then finds
\begin{equation}
    \xi_3(\gamma_{\opt})= \frac{\sigma_{pp,\opt}}{\left(\frac32\right)^{5/3}\gamma_{\opt}^{2/3}}=\frac{2^{1/3}}{3\lambda^{2/3}}\simeq \frac{2^{1/3}}{3}\left(\frac{c}{\mathcal N_\lambda-1}\right)^{4/3}\simeq 0.28
    (\mathcal N_\lambda-1)^{-4/3}.
\end{equation}
 One concludes that for large negativity the variance of $(\hat p+3\gamma \hat q^2)$ for optimal cubic Gaussian states is considerably smaller than the best such variance that can be obtained with a Gaussian state. For the values of $\lambda$, and hence of $\mathcal N_\lambda$, typically encountered in the literature (see Table~\ref{table:ideal}), $\xi_3(\gamma_{\opt})$ is well below $1$, so that the nonlinear squeezing constitutes an efficient non-gaussianity witness for those states. 

\begin{figure}\label{fig:wigneridealappr}
\centering 
  \subfloat[\label{fig:ferrini}][(a)]{\includegraphics[width=4cm]{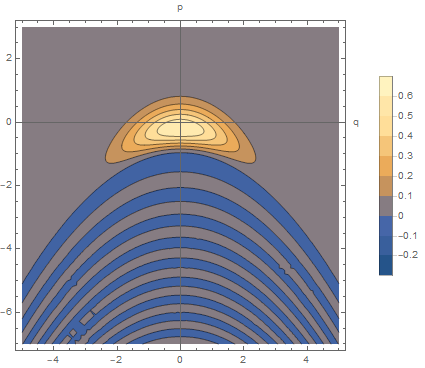}}
  \hspace{0.1cm}
   \subfloat[\label{fig:ErikssonNature}][(b)]{\includegraphics[width=4cm]{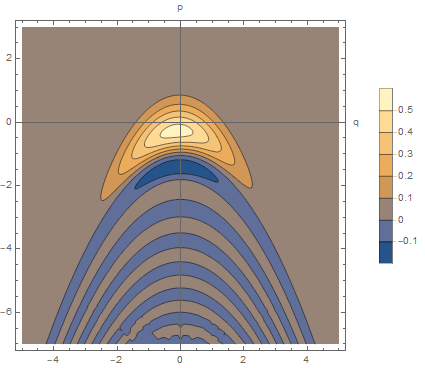}}
  \hspace{0.1cm}
   \subfloat[\label{fig:Yanagimoto}][(c)]{\includegraphics[width=4cm]{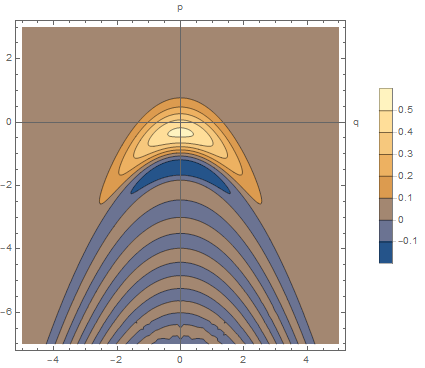}}
  \hspace{0.1cm}
  \subfloat[\label{fig:KalaFilipMarek}][(e)]{\includegraphics[width=4cm]{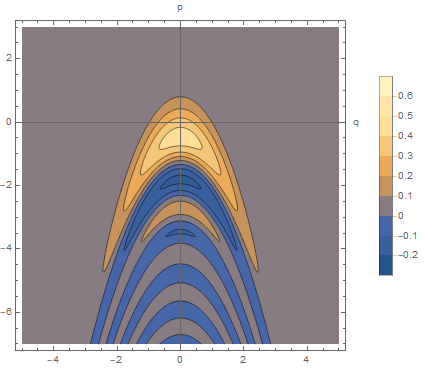}}
  \hspace{0.1cm}
   \caption{{Wigner function of  cubic squeezed states proposed in the literature (\cite{zheng_gaussian_2021}, \cite{eriksson_universal_2023}, \cite{yanagimoto_engineering_2020}, \cite{kala_cubic_2022}) as approximations of the ideal cubic state. As labelled, the parameters values for each figure are displayed in Table \ref{table:ideal} in the lines corresponding to the lables (a), (b), (c) and (e). Note that for (b), the Wigner function is displayed with a displacement of $\alpha = -0.090 + i0.025$.}
   }
   \label{fig:cubicphasestate_examples}
\end{figure}

We finally consider, for comparison, various  cubic squeezed vacuum states that have been proposed in the literature, experimentally or theoretically.  The Wigner functions of some of them (see~\cite{zheng_gaussian_2021,eriksson_universal_2023,yanagimoto_engineering_2020,kala_cubic_2022})  are plotted in Fig. \ref{fig:cubicphasestate_examples} and their Wigner negativities and QCS are reproduced in the left half of Table~\ref{table:ideal}. One observes that the negativities of these cubic Gaussian states are comparable to those of Fock states with $1,2,$ or at most $3$ photons {(see Table \ref{table:lambdan})} and that they have comparable QCS - be it slightly higher - than those of the latter. 
The second half of Table~\ref{table:ideal} shows the parameters of the optimal cubic Gaussian states with the same negativity. One notes in each case that the value of the QCS is about 5-10\% smaller, resulting in a similar increase in the decoherence time. Whether such optimization can be performed experimentally will depend on the range of squeezing and cubicity that are accessible in the experiment.

\captionsetup[subfloat]{labelformat=empty}



\section{Conclusion and outlook}\label{sec:conclusion}

We have, in this paper, made an exhaustive study of the cubic Gaussian states and identified, among them, the optimal cubic Gaussian states, defined to be those that, at fixed Wigner negativity, enjoy the longest coherence time (and hence the shortest quadrature coherence scale). We established that, nevertheless, this coherence time decreases quadratically in the Wigner negativity of the optimal cubic Gaussian states, showing that there is a necessary tradeoff between a large Wigner negativity and a short coherence time.  We have also investigated the sensitivity to thermal noise of the Wigner negativity and quadrature coherence scale of the optimal cubic Gaussian states and found that both quantities  are very sensitive to such noise.

We observed that a similar tradeoff between coherence time and Wigner negativity exists for Fock states. This leads us to conjecture  the persistence of such a relation in other families of states, such as for example, the nonlinearly squeezed vacuum states, that have attracted considerable experimental and theoretical attention~\cite{chang_2020, eriksson_2024, srinivas, fisher_1984, ashhab_2026, Fischer_2026}.
The explicit (asymptotic) formulas used in our analysis here are not available for those states and their analysis will henceforth be more challenging, but initial numerical investigations indicate that they have similar features. 
More generally, we formulate the following conjecture. Let 
\begin{equation}
\mathcal C(N)=\min_{\rho, \mathcal N(\rho)=N} \mathcal C(\rho) ;
\end{equation}
this is the smallest quadrature coherence scale (and hence the largest coherence time) of any state with a fixed Wigner negativity $N$. Then
\begin{equation}
\lim_{N\to+\infty} \mathcal C(N)=+\infty.
\end{equation}
More precisely, one may expect that, as for the Fock states and the optimal cubic Gaussian states
\begin{equation}
\mathcal C(N)\sim N,
\end{equation}
for large $N$. Establishing these conjectures on further subfamilies of states would already constitute an interesting progress and shed further light on the physical and operational meaning of Wigner negativity. It would  in particular establish that a large Wigner negativity always leads to a short decoherence time.

\medskip
\noindent{\it Acknowledgment}\\
This work was supported in part by the CNRS through the MITI interdisciplinary programs. M. Arnhem, S.~De~Bièvre, and G.~Patera acknowledge the support of the CDP C2EMPI, as well as the French State under the France-2030 programme, the University of Lille, the Initiative of Excellence of the University of Lille, the European Metropolis of Lille for their funding and support of the R-CDP-24-004-C2EMPI project.
The authors thanks Caroline Lasser, Virginia D'Auria, Valentina Parigi and Vojtech Kala for useful discussions.
GP acknowledge funding from the Plan France 2030 through the
projects OQuLus (ANR-22-PETQ-0013) and Gandalf (ANR-25-CE47-1818).
 
\newpage

\appendix


\section{Computation of the Wigner function of cubic Gaussian states}\label{app:wignercubic}
We derive here  Eq.~\eqref{eq:WignerLambda} and Eq.~\eqref{eq:wignercubicgaussianstate}. 
The Wigner function of $\hat\rho_{\textrm{G},\gamma}$ can be computed from Eq.~\eqref{eq:Winvkernel}:
\begin{eqnarray}\label{eq:A1}
    W_{\textrm{G},\gamma}(q,p)&=& \frac{1}{\pi}\int \exp(ipa)\exp\left(i\gamma((q+\frac{a}2)^3-(q-\frac{a}2)^3)\right)\langle q-\frac{a}{2}| {\rhoG}|q+\frac{a}{2}\rangle\rd a\nonumber ,\\
    &=&\frac{1}{\pi}\int \exp(ipa)\exp\left(i\gamma
    (3q^2+\frac{a^2}{4})a\right)
    \langle q-\frac{a}{2}|\rhoG|q+\frac{a}{2}\rangle\rd a. \label{eq:defwignergammaq}
\end{eqnarray}
Using Eq.~\eqref{eq:kernelsW} and Eq.~\eqref{eq:WignerGaussian}, one has 
\begin{equation}\label{eq:A2}
    \begin{split}
 \bra{q - \frac{a}{2}} \rhoG\ket{q + \frac{a}{2}} &= \frac12\int \exp(-i a p)\WG (q,p)\rd p, \\
 &= \sqrt{\frac{1}{2 \pi \sigma_{qq}}} \exp{\left[-2 \frac{\sigma_{pp}}{\det V}  (q-q_0)^2 \right]} \exp{ \left[- 4 \frac{\sigma_{qp}}{\det V} (q-q_0)p_0  \right]} \exp{  \left[-2 \frac{\sigma_{qq}}{\det V} p_0^2 \right]} \\
 & ~~~ \exp{\left[ \frac{ \left(4 \frac{\sigma_{qq}}{\det V} p_0 + 4 \frac{\sigma_{qp}}{\det V} (q-q_0) - i a \right)^2}{8 \frac{\sigma_{qq}}{\det V}} \right]},
 \end{split}
\end{equation}
where $\sigma'_{qq}$ etc. are defined in Eq.~\eqref{eq:definitionvariablesnew}.
Using Eq~\eqref{eq:A2} in Eq.~\eqref{eq:A1}, one finds
\begin{equation}
    \begin{split}
        \WGgamma(q,p) &= \frac{1}{\pi}\sqrt{ \frac{1}{2 \pi \sigma_{qq}}} \exp{\left[- 2 \left(\frac{\sigma_{pp}}\det V - \frac{\sigma^2_{qp} }{ \det V \sigma_{qq}} \right) (q-q_0)^2 \right]} \\
      & ~~~ {\int \exp{ \left[i \left((p - p_0) + 3 \gamma q^2 - \frac{\sigma_{qp}}{\sigma_{qq}} (q - q_0)  \right) a \right] }  \exp{ \left[- \frac{\det V}{8 \sigma_{qq}} a^2 \right] } \exp{ \left[i  \frac{\gamma }{4} a^3 \right] } \rd a.}
 \end{split}
\end{equation}
Using  the change of variables
\begin{equation}
\begin{split}
    s = \frac{a}{2} \frac{\det V}{\sqrt{\sigma_{qq}}}, \quad \lambda = 6 \gamma \left(\frac{\sigma_{qq}}{\det V}\right)^{3/2}, \quad x = 2\sqrt{\frac{\sigma_{qq}}{\det V}} \left( (p-p_0) + 3 \gamma q^2 - \frac{\sigma_{qp}}{\sigma_{qq}} (q-q_0) \right),
    \end{split}
\end{equation}
and recalling that 
$\frac{1}{\det V} \left(\sigma_{pp} - \frac{\sigma^2_{qp} }{ \sigma_{qq}}\right) = \frac{1}{4 \sigma_{qq}}$, one finds Eq.~\eqref{eq:WignerLambda}:
\begin{equation}\label{eq:A5}
     \WGgamma(q,p)  =  \frac{1}{\pi}\sqrt{ \frac{2}{ \pi \det{V}}} \exp{ \left[- \frac{1}{ 2\sigma_{qq}} (q-q_0)^2 \right]} \Lambda(x,\lambda), \end{equation}
where, for all $x,\lambda\in\R$,
\begin{equation}\label{eq:A6}
\Lambda(x,\lambda):= \int \rd s \exp(ixs-\frac12 s^2+i\lambda\frac{s^3}{3}).
 \end{equation}

 To derive Eq.~\eqref{eq:wignercubicgaussianstate}, we express $\Lambda(x,\lambda)$ in terms of the Airy function, as follows. 
Changing the integration variable to $y=\lambda^{1/3} s$, we find :
 \begin{equation}
     \Lambda(x,\lambda):= |\lambda|^{-1/3} \int \rd y \exp(i \lambda^{-1/3} xy-\frac12 \lambda^{-2/3} y^2+i\frac{y^3}{3}).
 \end{equation}
 Hence, the integral is now of the form $(\alpha>0)$ 
 \begin{equation}
 \textrm A(r,\alpha) = \int \rd y \exp(iry-\frac12 \alpha y^2+i\frac{y^3}{3})=\int \rd y \exp(i(r + \frac{\alpha^2}{4}) y+\frac{i}{3} (y + i \frac{\alpha}{2})^3-\frac{\alpha^3}{24}),
 \end{equation}
 with $r=\lambda^{-1/3}, \alpha=\lambda^{-2/3}>0$. 
With a second change of variables $t = y + i \frac{\alpha}{2}$, we then find
 \begin{equation}
 \begin{split}
      \textrm A(r, \alpha) & = \exp\left({\frac{\alpha r}{2}}\right) \exp\left({\frac{\alpha^3}{12}}\right) \int \rd t \exp\left(i(r + \frac{\alpha^2}{4}) t+\frac{i}{3} t^3\right), \\
     & = 2 \pi \exp\left({\frac{\alpha r}{2}}\right) \exp\left({\frac{\alpha^3}{12}}\right)  \textrm{Ai}(r + \frac{\alpha^2}{4}),
 \end{split}
 \end{equation}
 which yields~\cite{brunelli_linear_2019} 
\begin{equation}\label{eq:BrunelliHouhouFormula}
\Lambda(x,\lambda)=2\pi|\lambda|^{-1/3}\exp(\frac{1}{12\lambda^2})\exp(\frac{x}{2\lambda}) \textrm{Ai}(\lambda^{-1/3}(x+\frac1{4\lambda})).
 \end{equation}
This, together with Eq.~\eqref{eq:A5}, implies Eq.~\eqref{eq:wignercubicgaussianstate}.

It is clear from its definition in Eq.~\eqref{eq:A6}  that $\Lambda(x,\lambda)$ is, for every $x\in\R$, an analytic function of $\lambda$. In particular, 
\begin{equation}
\Lambda(x,0)=\lim_{\lambda\to0}\Lambda(x,\lambda)=\sqrt{2\pi}\exp(-\frac12x^2).
\end{equation}
Let us point out that this limit is less obvious from the expression in the right hand side of Eq.~\eqref{eq:BrunelliHouhouFormula} which contains terms that are singular at $\lambda=0$. 

Note finally that $\Lambda(x,\lambda)=0$ iff 
$$
x=x_n:=-\frac1{4\lambda}+\lambda^{1/3}\eta_n<0,
$$
where $\eta_n$ is a zero of the Airy function. 

\section{Invariance of the Wigner negativity under displacement}\label{app:wignernegdisplinvariance}
We show here that the Wigner negativity of the translated state $\hat\rho$,  $\hat{\rho}_{\gamma, \alpha} = \hat{\Gamma}_\gamma \hat{D}(\alpha) \rhoh \hatd{D}(\alpha) \hatd{\Gamma}_\gamma$ where $\alpha = \frac{1}{\sqrt{2}}(q_0 + i p_0)$, does not depend on $\alpha$. 
For that purpose, note that one can commute the displacement operator and the cubic phase gate operator as follows:
\begin{equation}
    \rhoh_{\gamma, \alpha} = \hat D(\alpha) \exp{(- i \gamma K(\hat q))} \rhoh_{\gamma}  
 \exp{(+ i \gamma K(\hat q))}  \hatd D(\alpha),
\end{equation}
where $K(\hat q) =  3 q_0 \hatn{q}{2} + 3 q_0^2 \hat q + q_0^3$ is a self-adjoint operator quadratic in $\hat q$. Hence $\exp{(- i \gamma K(\hat q))}$ is a Gaussian unitary and, writing $W_{\gamma,\alpha}$ for the Wigner function of $\hat\rho_{\gamma, \alpha}$, one has 
\begin{equation}\label{eq:Wignercubicdispl}
    W_{\gamma, \alpha}(q,p) = W_\gamma (\Phi(q, p)),
\end{equation}
where $\Phi$ is the linear symplectic transformation generated by the Hamiltonian $K(q)=  3 q_0 q^2 + 3 q_0^2 q + q_0^3$, such that $\Phi(q, p) = (q , p - 6 \gamma q_0 q - 3 \gamma q_0^2)$.
It follows that $\rhoh_{\gamma, \alpha}$ has the same negativity as $\rhoh_{\gamma}$:
\begin{equation}\label{eq:wignernegcubicdispl}
    \mathcal{N}_{\gamma, \alpha} =\frac12 \int \norm{W_{\gamma,\alpha} (q,p)} dq dp =\frac12 \int \norm{W_{\gamma} (\Phi(q,p))} dq dp =\frac12 \int \norm{W_{\gamma} (q,p)} dq dp = \mathcal{N}_\gamma.
\end{equation}
In conclusion, whereas the Wigner function of $W_{\gamma, \alpha}$ does depend on $\alpha$, its negativity does not.

\section{Mathematical analysis of the Wigner negativity of cubic Gaussian states}\label{app:bounds}
In this appendix we show Eq.~\eqref{eq:wignerneg} and the bounds Eq.~\eqref{eq:wignernegupperboundsmallgamma} and~\eqref{eq:wignernegupperboundlargegamma}.

Starting from Eq.~\eqref{eq:wignercubicgaussianstate}, and performing the change of variables $(q,p)\to (q,\eta)$, one finds:
\begin{equation}
    \begin{split}
        \mathcal N(\rhoGgamma)&=\frac{1}{2}  \iint |W_{G,\gamma}(q,p)|\rd q\rd p, \\
        & = \sqrt{ \frac{2}{ \pi \det{V}}}  |\lambda|^{-1/3}\exp(-\frac{1}{24\lambda^2}) \iint \exp{ \left[- 2 \left(\sigma'_{pp} - \frac{\sigma'^2_{qp} }{ \sigma'_{qq}} \right) (q-q_0)^2 \right]} \rd q \exp(\frac{\eta}{2\lambda^{2/3}}) \textrm{Ai}(\eta)) \rd p, \\
         & = \frac{1}{2 \sqrt{\sigma'_{qq}}}\sqrt{ \frac{2}{ \pi \sigma_{qq}}} |\lambda|^{-1/3} \lambda^{1/3}  \exp(-\frac{1}{24\lambda^2}) \iint \exp{ \left[- 2 \left(\sigma'_{pp} - \frac{\sigma'^2_{qp} }{ \sigma'_{qq}} \right) (q-q_0)^2 \right]} \rd q \exp(\frac{\eta}{2\lambda^{2/3}}) \textrm{Ai}(\eta) \rd \eta, \\
                 & =  \sqrt{ \frac{\det{V}}{2 (\sigma_{qq} \sigma_{pp} -\sigma^2_{qp}  )}}  \exp(-\frac{1}{24\lambda^2}) \int \exp(\frac{\eta}{2\lambda^{2/3}}) \textrm{Ai}(\eta) \rd \eta, \\
                 & = \exp(-\frac{1}{24\lambda^2}) \int \exp(\frac{\eta}{2\lambda^{2/3}}) \textrm{Ai}(\eta) \rd \eta, \\
    \end{split}
\end{equation}
which is Eq.~\eqref{eq:wignerneg}. 

Thinking of $\eta$, defined in {Eq.~\eqref{eq:etadef} together with Eqs.~\eqref{eq:lambdadef} and \eqref{eq:xdef}} as a function on phase space, we designate by $\{\eta< \eta_0\}$ the region where $\eta$ is smaller than the largest (negative) zero $\eta_0$ of the Airy function. One then has 
\begin{equation}
R_-:=\{(q,p)|W(\rhoGgamma)(q,p)<0\}\subset \{\eta<\eta_0\}.
\end{equation}
Consequently, using the same change of variables as above, one finds 
\begin{equation}
    \mathcal N(\rhoGgamma)=1+2\int_{R_-} |W|(\rhoGgamma)(q,p)\rd q\rd p\leq 1+2J^-_\lambda,
\end{equation}
where
\begin{equation}\label{eq:Jlambdamin}
 J_\lambda^- :=\iint_{\eta<\eta_0} |W(\rhoGgamma)(q,p)|\rd q\rd p=\exp(-\frac1{24\lambda^2}) \int_{-\infty}^{\eta_0} \exp(\frac{\eta}{2\lambda^{2/3}})|\textrm{Ai}(\eta)| \rd\eta.
\end{equation}
We will prove two estimates valid for all $\lambda$, one being more efficient for small $\lambda$, the other for large $\lambda$. The Airy function is bounded so we have
\begin{eqnarray}\label{eq:Jgammamin}
J_\lambda^-
&\leq& M_0\exp(-\frac1{24\lambda^2}) \int_{-\infty}^{\eta_0} \exp(\frac{\eta}{2\lambda^{2/3}}) \rd\eta\nonumber\\
& =  & 2 M_0\lambda^{2/3}\exp(-\frac1{24\lambda^2})\exp(-\frac{|\eta_0|}{2\lambda^{2/3}}),
\end{eqnarray}
where 
\begin{equation}
M_0=\max_{\eta<\eta_0} |\textrm{Ai}(\eta)|.
\end{equation}
The maximum $M_0$ can be evaluated numerically:
\begin{equation}
M_0\approx 0.42.
\end{equation}
This ends the proof of Eq.~\eqref{eq:wignernegupperboundsmallgamma}.

For large $\lambda$, the integral over $\eta$ undergoes an effectif cutoff at large negative $\eta$ of size $\eta\simeq -\lambda^{2/3}$; the above estimate is then far from optimal. Instead, we write
\begin{eqnarray*}
J_\lambda^-&=&\exp(-\frac1{24\lambda^2}) \int_{-\infty}^{\eta_0} \exp(\frac{\eta}{2\lambda^{2/3}})|\textrm{Ai}(\eta)| \rd\eta,\\
&=&\exp(-\frac1{24\lambda^2})\lambda^{2/3} \int_{-\infty}^{\eta_0\lambda^{2/3}} \exp(\frac{\sigma}{2})|\textrm{Ai}(\lambda^{2/3}\sigma)| \rd\sigma.
\end{eqnarray*}
where we made the change of variables $\sigma=\eta/\lambda^{2/3}$.
Provided $\lambda>> 1$, we can then use the asymptotic expansion of the Airy function
\begin{equation}\label{eq:airyasympneg}
\textrm{Ai}(x)\sim \frac{\sin(\frac23|x|^{3/2}+\frac{\pi}{4})}{\sqrt\pi x^{1/4}}, x\to-\infty,
\end{equation}
 to obtain the bound 
\begin{eqnarray}
   J_\lambda^-&\leq& \exp(-\frac1{24\lambda^2})\lambda^{1/2}\frac{1}{\sqrt\pi} \int_{-\infty}^{\eta_0\lambda^{2/3}} \exp(\frac{\sigma}{2})| \frac{\rd\sigma}{|\sigma|^{1/4}}, \\
   &\leq & \lambda^{1/2}\exp(-\frac1{24\lambda^2})\frac{1}{\sqrt\pi} \int_{0}^{+\infty} \exp(-\frac{\sigma}{2})\frac{\rd\sigma}{\sigma^{1/4}}, \\
   & = & \frac{2^{3/4}}{\sqrt\pi} \Gamma(\frac{3}{4}) \lambda^{1/2}\exp(-\frac1{24\lambda^2}) . 
\end{eqnarray}
This then proves Eq.~\eqref{eq:wignernegupperboundlargegamma}.


\section{Computing the QCS of cubic Gaussian states}\label{app:QCScubic}
Overlap formula:
\begin{equation}\label{eq:overlapformula}
    \tr (\hat{F}_1 \hat{F}_1) = \frac{\pi}{2} \iint W_{F_1}(q,p) W_{F_2}(q,p) \rd q \rd p.
\end{equation}
We need the Wigner function for $\rhoG q^2$ and $q^2 \rhoG$. They are the complex conjugate of each other. Let us start with $q^2 \rhoG$:
\begin{equation}
\begin{split}
    W_{\hat{q}^2 \rhoG} & = \frac{1}{2 \pi} \int \exp{(i p x)} \bra{q - \frac{x}{2}} \hat{q}^2 \rhoG \ket{q + \frac{x}{2}} \rd x, \\
    & =   \frac{1}{2 \pi} \int \exp{(i p x)} \left(q - \frac{x}{2}\right)^2 \bra{q - \frac{x}{2}}  \rhoG \ket{q + \frac{x}{2}} \rd x, \\
        & =   \frac{1}{2 \pi}  \left(q + \frac{i}{2} \frac{\partial}{\partial p}\right)^2 \int \exp{(i p x)} \bra{q - \frac{x}{2}}  \rhoG \ket{q + \frac{x}{2}} \rd x, \\
    & =  \left(q + \frac{i}{2} \frac{\partial}{ \partial p}\right)^2  W_{\rhoG}(q,p),
\end{split}
\end{equation}
such that, as $W_{\rhoG \hat{q}^2} = W_{\hat{q}^2 \rhoG}^*$, we find that:
\begin{equation}
    W_{[\rhoG,q^2]}(q,p) =  W_{\rhoG q^2}(q,p) -  W_{q^2 \rhoG}(q,p) = - 2 i q \frac{\partial}{\partial p} W_{\rhoG} (q,p),
\end{equation}
where $W_{\rhoG} (q,p)$ is the Wigner function associated to the state $\rho_G$. As we will deal with Gaussian states, we identify $W_{\rhoG} (q,p) = W_{G} (q,p)$.
By making use of the overlap formula Eq.(\ref{eq:overlapformula}), we find for centered Gaussian states $\hat\rho_{G^0}$ that:
\begin{equation}
    \begin{split}
       \tr[\hat\rho_{G^0},\hat q^2][\hat q^2,\hat\rho_{G^0}] & = \frac{\pi}{2} \iint  W_{[\hat\rho_{G^0},q^2]}(q,p) W_{[q^2,\hat\rho_{G^0}]}(q,p) \rd q \rd p
   = 4 \frac{\sigma_{qq}^2}{(\det V)^{3/2}}.
    \end{split}
\end{equation}
Similarly, we can compute 
\begin{eqnarray}\label{eq:tracecommutatorq1}
   \tr[\hat\rho_{G^0},\hat q][\hat q,\hat\rho_{G^0}] &=&   \frac{\pi}{2} \iint \left(\frac{\partial }{\partial p} W_{G^0}(q,p)\right)^2 \rd q \rd p
   = 2 \frac{\sigma_{qq}}{(\det V)^{3/2}},
\end{eqnarray}
\begin{eqnarray}\label{eq:tracecommutatorpq}
   \tr[\hat\rho_{G^0},\hat p][\hat q,\hat\rho_{G^0}] &=&   \frac{\pi}{2} \iint (\frac{\partial }{\partial p}  W_{G^0}(q,p)) (\frac{\partial }{\partial q} W_{G^0}(q,p)) \rd q \rd p 
   = -2 \frac{\sigma_{qp}}{(\det V)^{3/2}}.
\end{eqnarray}
\newline
Hence, we can express the QCS for a cubic Gaussian state from Eq.(\ref{eq:QCScubicdispl}) in terms of the covariance matrix elements as 
\begin{equation}\label{eq:QCScubicsigma}
    \begin{split}
         \mathcal C^2_{\textrm G, \gamma}
           &=\frac{1}{\det V} \left( \sigma_{qq} + \sigma_{pp} + 18 \gamma^2 \sigma_{qq}^2  + 12 \gamma q_0 \left( 3 \gamma q_0 \sigma_{qq} + \sigma_{qp} \right) \right),
    \end{split}
\end{equation}
which is equal to the expression of Eq.(\ref{eq:QCScubiccovelements}).
{We can express it in terms of the most general parametrization of the Gaussian states 
:
{\begin{eqnarray}
    \frac{\sigma_{qq}}{\det V} &=& \left(\frac{1-q}{1+q}\right) \frac{(\cosh(2r) - \cos(\phi) \sinh(2r))}{2}, \\
    \frac{\sigma_{pp}}{\det V} &=& \left(\frac{1-q}{1+q}\right) \frac{(\cosh(2r) + \cos(\phi) \sinh(2r))}{2}, \\
     \frac{\sigma_{qq} + \sigma_{pp}}{\det V} &=& \left(\frac{1-q}{1+q}\right) \cosh(2r), \\
     \frac{\sigma_{qp}}{\det V} &=& - \left(\frac{1-q}{1+q}\right) \frac{(\sinh(2r)\sin(\phi))}{2} , \\
     \frac{\sigma_{qq}^2}{\det V} &=& \frac{1}{4}(\cosh(2r) - \cos(\phi) \sinh(2r))^2.
\end{eqnarray}}
Hence,
\begin{equation}
\begin{split}
    \mathcal C^2_{\textrm G, \gamma, r} 
        =& \left(\frac{1-q}{1+q}\right) \left[ \cosh(2r) + 18 \gamma^2 q_0^{2} (\cosh(2 r) - \cos(\phi) \sinh(2 r)) -6 \gamma q_0  \sinh(2r) \sin(\phi) \right]\\
        & + \frac92 \gamma^2 (\cosh(2 r) - \cos(\phi) \sinh(2 r))^2,
\end{split} 
\end{equation}
}
\newline
{This is the most general expression of the QCS in terms of the 6 Gaussian parameters and the cubicity of the cubic phase state. Interestingly, this expression is independent of the displacement in the p-direction but depends on the displacement in the q-direction in phase space through the presence $q_0$ in the above equation. Hence, one can ask what is the value of $q_0$ for which the QCS is minimal. The answer is :
\begin{equation}\label{eq:q0min}
    q_0^{min} = - \frac{\sigma_{qp}}{6 \gamma \sigma_{qq}} = - \frac{1}{6 \gamma} \frac{\sinh(2 r) \sin(\phi)}{(\cosh(2 r) - \cos(\phi) \sinh(2 r))}.
\end{equation}
Hence, in the absence of off-diagonal terms $\sigma_{qp}$ in the covariance matrix of Gaussian states (which means that its wigner function is squeezed along the q- or p-axis), the QCS is minimal when the state is not displaced along the q-axis. Note that the global minimum of the QCS is reached for $\phi=\pi$, i.e. for a p-squeezed state. However, for rotated states (when $\sigma_{qp} \neq 0$), then the QCS can be locally minimized by displacing the initial Gaussian state by $q_0^{min}$. Hence,
\begin{equation}\label{eq:QCSgammaq0min}
    \begin{split}
        C^2_{G, \gamma, q_0^{min}} & = \frac{1}{\det V} \left( \sigma_{qq} + \sigma_{pp} + 18 \gamma^2 \sigma_{qq}^2 - \frac{\sigma_{qp}^2}{\sigma_{qq}}\right), \\
        & = \left(\frac{1-q}{1+q}\right) \left[ \cosh(2 r) + \frac{9}{2} \gamma^2 \left(\frac{1+q}{1-q}\right) (\cosh(2 r) - \cos(\phi) \sinh(2 r))^2 - \frac{\sinh(2r)^2 \sin(\phi)^2}{2 (\cosh(2 r) - \cos(\phi) \sinh(2 r))} \right].
    \end{split}
\end{equation}
}

\section{Minimization of the QCS}
By replacing the cubicity $\gamma$ by the parameter $\lambda$ defined in Eq.\eqref{eq:lambdadef} in the QCS expression of Eq.\eqref{eq:QCScubicsigma}, we find :
\begin{equation}
\begin{split}
    C^2_{G,\lambda} 
    & = \frac{\sigma_{qq} + \sigma_{pp}}{\det V} + \lambda^2 \left(\frac{\det V}{\sigma_{qq}}\right)^{2}\left(\frac{\sigma_{qq}}{2} + q_o^2 \right) + 2 \lambda q_0 \left(\frac{\det V}{\sigma_{qq}}\right)^{1/2} \frac{\sigma_{qp}}{\sigma_{qq}}.
\end{split}
\end{equation}
Hence, one can minimize the QSC $ C^2_{G,\lambda} $ over the displacement $q_0$  ? Indeed, the optimal value of $q_0^{\opt}$ is written in terms of $\lambda$ and the elements of the covariance matrix as 
\begin{equation}
    q_0^{\opt} = - \frac{1}{\lambda} \left(\frac{\sigma_{qq}}{\det V}\right)^{3/2} \frac{\sigma_{qp}}{\sigma_{qq}},
\end{equation}
such that this optimal displacement is proportionnal to the off-diagonal elements $\sigma_{qp}$ in the covariance matrix of the kernel Gaussian states $\rho_G$.
Hence, 
\begin{equation}
\begin{split}
    C^2_{G,\lambda,q_0^{\opt}} 
     & = \frac{\sigma_{qq} + \sigma_{pp}}{\det V} + \frac{\lambda^2}{2} \frac{(\det V)^2}{\sigma_{qq}} - \frac{\sigma_{qp}}{\det V} \frac{\sigma_{qp}}{\sigma_{qq}}, \\
     & = \left(\frac{1-q}{1+q}\right) \left( \cosh(2r) + \frac{1}{\cosh(2r) - \cos(\phi) \sinh(2r)} \left[ \lambda^2 \left(\frac{1+q}{1-q}\right)^4 - \frac{\sinh(2r)^2 \sin(\phi)^2}{2} \right] \right). \\
\end{split}
\end{equation}
Now, by taking $\phi = \phi_{\opt} = \pi$, we find :
\begin{equation}
      C^2_{G,\lambda,q_0^{\opt},\phi_{\opt}} = \left(\frac{1-q}{1+q}\right) \left( \cosh(2r) + e^{- 2 r} \lambda^2 \left(\frac{1+q}{1-q}\right)^4 \right),
\end{equation}
in agreement with Eq. (\ref{eq:QCSGaussianCentered}) for $\phi=\pi$ as expected.


\section{Computation of the Mandel parameter for cubic squeezed vacuum states} \label{sec:mandel}

\subsection{Case of the cubic vacuum state}

In this section we compute the Mandel Q parameter for the cubic vacuum state $\Gamma_\gamma \ket{0}$. As a reminder, the Mandel Q parameter can be written as :
\begin{equation}
    Q = \frac{\mean{\hatdn{a}{2} \hatn{a}{2}} - \mean{\hatd{a} \hat{a}}^2}{\mean{\hatd{a} \hat{a}}}.
\end{equation}
It is larger than zero for classical states. Hence, a negative value of the Mandel Q parameter is a sufficient condition of nonclassicality.
In order to compute the value of the Mandel Q parameter for the cubic vacuum states, we need to compute the following :
\begin{eqnarray}
    \hat{a} \hat{\Gamma}_\gamma \ket{0} 
    &=& \left(\frac{-3 i \gamma}{2 \sqrt{2}}\right) \hat{\Gamma}_\gamma  \left( \ket{0} + \sqrt{2} \ket{2} \right),
\end{eqnarray}
such that, as $\hat{\Gamma}_\gamma$ is a unitary operator,
\begin{equation}
    \mean{\hatd{a} \hat{a}}_\gamma  
    = \frac{27}{8} \gamma^2.
\end{equation}
Similarly, we can find $\mean{\hatdn{a}{2} \hatn{a}{2}}_\gamma$ by computing
\begin{eqnarray}
    \hatn{a}{2} \hat{\Gamma}_\gamma \ket{0} 
        &=& \hat{\Gamma}_\gamma \left(\frac{-3 i \gamma}{2 \sqrt{2}}\right) \left(\left(\frac{-3 i \gamma}{2 \sqrt{2}}\right)  \left(3 \ket{0} + 6 \sqrt{2} \ket{2} + 2 \sqrt{6} \ket{4}\right) + 2 \ket{1} \right) ,\\
\end{eqnarray}
such that 
\begin{eqnarray}
    \mean{\hatdn{a}{2} \hatn{a}{2}}_\gamma 
    &=& \frac{9}{64} \gamma^2 \left( 32 + 945 \gamma^2 \right).
\end{eqnarray}
Hence, we can compute the Mandel Q parameter 
\begin{equation}
    Q = \frac{4}{3} + 36 \gamma^2.
\end{equation}
This result might look surprising, as for $\gamma=0$, hence the vacuum state, the Mandel Q parameter is equal to $4/3$ which seems arbitrary. However, let us keep in mind that Mandel Q parameter for vacuum state is undefined (as it involves a division by 0).
Nevertheless, this result shows that the Mandel parameter never detects the nonclassicality of the cubic vacuum state. Hence, other nonclassicality witnesses are needed such as the Wigner Negativity.
\subsection{Case of the cubic squeezed state}

We now extend the above computation to cubic squeezed vacuum states. Recall that 
\begin{equation}
    \hatd{S}_{r, \phi} \hat{a}  \hat{S}_{r, \phi} = \cosh{(r)} \hat{a} - e^{i \phi} \sinh{(r)} \hatd{a}.
\end{equation}
Hence, the action of the annihilation operator on the cubic squeezed state is:
\begin{eqnarray*}
    \hat{a} \hat{\Gamma}_\gamma \hat{S}_{r, \pi} \ket{0} 
    &=& \hat{\Gamma}_\gamma \hat{S}_{r, \pi} \left(\left(\frac{-3 i \gamma e^{2 r}}{2 \sqrt{2}}\right)   \left( \ket{0} + \sqrt{2} \ket{2} \right) + \sinh{(r)} \ket{1} \right).
\end{eqnarray*}
The mean number  of a cubic squeezed state is therefore 
\begin{equation}
    \mean{\hatd{a} \hat{a}}_{\gamma, r, \pi} 
    = \frac{27}{8} \gamma^2 e^{4 r}  + \sinh{(r)}^2.
\end{equation}
In order to compute the fourth order moment involved in the Mandel parameter, we need to compute:
\begin{eqnarray*}
    \hat{a}^2 \hat{\Gamma}_\gamma \hat{S}_{r, \pi} \ket{0} &=& \hat{a} \left( \left[\hat{a},\hat{\Gamma}_\gamma \right] \hat{S}_{r, \pi} +  \hat{\Gamma}_\gamma \hat{a} \hat{S}_{r, \pi} \ket{0} \right), \\
     &=& \hat{\Gamma}_\gamma \hat{S}_{r, \pi} \left(\left(\frac{-3 i \gamma}{\sqrt{2}}\right) e^{2 r} \hatn{q}{2} + \cosh{(r)} \hat{a} + \sinh{(r)} \hatd{a} \right) \left(\left(\frac{-3 i \gamma e^{2 r}}{2 \sqrt{2}}\right)   \left( \ket{0} + \sqrt{2} \ket{2} \right) + \sinh{(r)} \ket{1} \right). 
\end{eqnarray*}
We introduce the operator $\hat{B} = \left(\left(\frac{-3 i \gamma}{\sqrt{2}}\right) e^{2 r} \hatn{q}{2} + \cosh{(r)} \hat{a} + \sinh{(r)} \hatd{a} \right) $ and analyse its action on the three Fock state $\ket{0}, \ket{1}$ and $\ket{2}$.
First, $\hat{B} \ket{0}$ has already been computed in the previous subsection :
\begin{eqnarray*}
  \left(\frac{-3 i \gamma}{2 \sqrt{2}}\right) e^{2 r}  \hat{B} \ket{0} 
     &=&  \left(\left(\frac{-9  \gamma^2 e^{4 r}}{8}\right)  \left(\ket{0} + \sqrt{2} \ket{2}\right) \right) + \left(\frac{-3 i \gamma}{2 \sqrt{2}}\right) e^{2 r} \sinh{(r)} \ket{1}, 
\end{eqnarray*}
Next, for $\hat{B}\ket{1}$, we find :
\begin{eqnarray*}
  \sinh{(r)}  \hat{B} \ket{1} 
    &=&  \cosh{(r)}  \sinh{(r)} \ket{0} + 3 \left(\frac{-3 i \gamma}{2 \sqrt{2}}\right) e^{2 r}  \sinh{(r)} \ket{1} + \sqrt{2} \sinh{(r)}^2 \ket{2} + \sqrt{6}  \left(\frac{-3 i \gamma}{2 \sqrt{2}}\right) e^{2 r}  \sinh{(r)} \ket{3}.
\end{eqnarray*}
And, finally, for  $\hat{B}\ket{2}$:
\begin{eqnarray*}
 \sqrt{2} \left(\frac{-3 i \gamma}{2 \sqrt{2}}\right) e^{2 r}  \hat{B} \ket{2} 
     &=&  2 \left(\frac{-9 \gamma^2 e^{4r}}{8}\right) \ket{0} + 2 \left(\frac{-3 i \gamma}{2 \sqrt{2}}\right) e^{2 r} \cosh{(r)} \ket{1} + 5 \sqrt{2} \left(\frac{-9 \gamma^2 e^{4 r}}{8}\right)  \ket{2} + \\
    && \sqrt{6} \left(\frac{-3 i \gamma}{2 \sqrt{2}}\right) e^{2 r} \sinh{(r)} \ket{3} + 2 \sqrt{6} \left(\frac{-9 \gamma^2 e^{4 r}}{8}\right)  \ket{4}.
\end{eqnarray*}
Adding up all terms, we find 
\begin{eqnarray}
    \hat{a}^2 \hat{\Gamma}_\gamma \hat{S}_{r, \pi} \ket{0} &=&\hat \Gamma_\gamma S_{r,\pi}\left(\sum_{n=0}^4 c_n|n\rangle\right),\label{eq:a2_cubic}
\end{eqnarray}
where
\begin{eqnarray*}
    c_{0} &=& \left(\frac{-3 i \gamma }{2 \sqrt{2}}\right)^2 e^{4 r} +  \cosh{(r)}  \sinh{(r)} + 2 \left(\frac{-3 i \gamma}{2 \sqrt{2}}\right)^2 e^{4 r} = 3 \left(\frac{-9 \gamma^2 }{8}\right) e^{4 r} +  \cosh{(r)}  \sinh{(r)}, \\
    c_{1} &=& \left(\frac{-3 i \gamma}{2 \sqrt{2}}\right) e^{2 r} \sinh{(r)} + 3 \left(\frac{-3 i \gamma}{2 \sqrt{2}}\right) e^{2 r}  \sinh{(r)} + 2 \left(\frac{-3 i \gamma}{2 \sqrt{2}}\right) e^{2 r} \cosh{(r)} = 2 \left(\frac{-3 i \gamma}{2 \sqrt{2}}\right) e^{2 r} \left( 2 \sinh{(r)}  + \cosh{(r)} \right), \\
    c_{2} &=& \sqrt{2} \left(\frac{-3 i \gamma }{2 \sqrt{2}}\right)^2 e^{4 r} + \sqrt{2} \sinh{(r)}^2 + 5 \sqrt{2}\left(\frac{-3 i \gamma}{2 \sqrt{2}}\right)^2 e^{4 r} = 6 \sqrt{2} \left(\frac{- 9 \gamma^2}{8}\right) e^{4 r} + \sqrt{2} \sinh^2{(r)} , \\
    c_{3} &=& \sqrt{6}  \left(\frac{-3 i \gamma}{2 \sqrt{2}}\right) e^{2 r}  \sinh{(r)} + \sqrt{6} \left(\frac{-3 i \gamma}{2 \sqrt{2}}\right) e^{2 r} \sinh{(r)} = 2 \sqrt{6} \left(\frac{-3 i \gamma}{2 \sqrt{2}}\right) e^{2 r} \sinh{(r)} , \\
    c_{4} &=& 2 \sqrt{6} \left(\frac{-9 \gamma^2}{8}\right) e^{4 r}.
\end{eqnarray*}
These coefficients reduce to the ones found in the previous section in the limit of $r=0$.
Their square moduli are:
\begin{eqnarray}
|c_0|^2 &=& 9 \left(\frac{9\gamma^2}{8} e^{4r}\right)^2
+ \cosh^2 (r)\, \sinh^2 (r) + 6 \left(-\frac{9\gamma^2}{8} e^{4r} \right)\cosh (r)\, \sinh (r)  , \\
|c_1|^2 &=& 4 \, \left(\frac{9\gamma^2 e^{4r}}{8}\right)
\bigl(\cosh^2 (r) + 4 \sinh^2 (r) + 4 \cosh (r) \sinh (r) \bigr),\\
|c_2|^2 &=&
72 \left(\frac{9\gamma^2 e^{4r}}{8} \right)^2
+ 2\sinh^4 (r) - 24 \left(\frac{9\gamma^2 e^{4r}}{8} \right) \sinh^2 (r) ,\\
|c_3|^2 &=&
24 \, \left(\frac{9\gamma^2 e^{4r}}{8}\right)\, \sinh^2 (r),\\
|c_4|^2 &=&
24 \, \left(\frac{9\gamma^2 e^{4r}}{8}\right)^2. 
\end{eqnarray}
Finally, we find
\begin{eqnarray*}
\mean{\hatdn{a}{2} \hatn{a}{2}} _{\gamma, r,\pi} = \sum_{k=0}^4 |c_k|^2 
&=& \frac{8505}{64} \gamma^{4} e^{8r}
+ \frac{9}{4} \gamma^{2} e^{4r} \left(2 + 10 \sinh^2 (r) + 5 \cosh (r) \sinh (r)\right)
+ \sinh^2(r)\left(\cosh^{2}(r)+2 \sinh^2(r)\right).
\end{eqnarray*}
Hence, the numerator of the Mandel Q parameter is :
\begin{eqnarray*}
    \mean{\hatdn{a}{2} \hatn{a}{2}} _{\gamma, r,\pi} - \mean{\hatd{a} \hat{a}}_{\gamma, r,\pi}^2 
&=& 96\, \frac{81}{64} e^{8r} \gamma^{4}
+ \frac{9}{4} e^{4r} \gamma^{2} \left(2 + 7 \sinh^2(r) + 5\cosh(r)\sinh(r)\right) + \sinh^2(r)\left(\cosh^{2}(r)+ \sinh^2(r)\right),
 \\
\end{eqnarray*}
and the Mandel Q parameter of the squeezed cubic states is found to be: 
\begin{eqnarray}
    Q_{\gamma, r, \pi} = \frac{\frac{7776}{8} e^{8r} \gamma^{4}
+ 18 \, e^{4r} \gamma^{2} \left(2 + 7 \sinh^2(r) + 5\cosh(r)\sinh(r)\right) + 8 \sinh^2(r)\left(\cosh^{2}(r)+ \sinh^2(r)\right)}{ 27 \gamma^2 e^{4 r}  + 8 \sinh{(r)}^2 }.
\end{eqnarray}
In the limit of $r=0$, we have : 
\begin{eqnarray}
    Q_{\gamma, r=0,} = \frac{96 \frac{81}{8}\gamma^{4}
+ 36 \, \gamma^{2}}{ 27 \gamma^2} = \frac{96 \,\frac{9}{8}\gamma^{2}
+ 4}{3} = 36 \gamma^{2}
+ \frac{4}{3} ,
\end{eqnarray}
which matches Eq.(\ref{eq:mandelcubicvacuum})

\bibliography{202607_main}

\end{document}